\def\C(#1){|#1|}

\def\eqar#1{\begin{eqnarray} #1 \end{eqnarray}}
\def\eq#1{\begin{equation} #1 \end{equation}}
\def\eqn#1{\begin{equation} \nonumber #1 \end{equation}}

\def\mbit#1{\mbox{\boldmath$ #1 $}}
\def\Def{\buildrel \Delta \over =}

\def\Ind(#1){\Delta\left[#1\right]}

\def\Inner(#1,#2){\langle #1, #2 \rangle}

\documentclass[journal, letterpaper, onecolumn,draftclsnofoot,12pt]{IEEEtran}
\usepackage{graphicx}
\usepackage{amsmath}
\usepackage{amsfonts}
\usepackage{array}
\newtheorem{thm}{Theorem}
\newtheorem{cor}[thm]{Corollary}

\newtheorem{obs}[thm]{Observation}

\newcounter{definition}
\newcounter{example}
\newcounter{theorem}
\newcounter{lemma}
\newcounter{corollary}
\newenvironment{example}
{\vspace{3mm}\refstepcounter{example} \noindent{\tt
Example~\theexample} \\}
 {\hfill$\diamond$\par\vskip1mm}

\newenvironment{theorem}
{\vspace{3mm}\refstepcounter{theorem} \noindent{\bf Theorem
\arabic{theorem}}\\}
{\hspace*{\fill}~$\bigtriangledown$\par\unskip}

\newenvironment{lemma}
{\vspace{3mm}\refstepcounter{lemma} \noindent{\bf Lemma
\arabic{lemma}}\\}
{\hspace*{\fill}~$\bigtriangledown$\par\unskip}

\newenvironment{corollary}
{\vspace{3mm}\refstepcounter{corollary} \noindent{\bf Corollary
\arabic{corollary}}\\}
{\hspace*{\fill}~$\bigtriangledown$\par\unskip}

\begin{document}
\title{On the Weight Enumerator and the Maximum Likelihood Performance of Linear Product Codes
\footnote{
This work was presented in part at the WirelessCom Symposium on Information Theory, Maui, Hawaii, USA in June 2005.}}

\author{Mostafa El-Khamy and Roberto Garello
\footnote{Mostafa El-Khamy is with the Department of Electrical
Engineering,
  California Institute of Technology, Pasadena, CA 91125, USA,
  email: mostafa@systems.caltech.edu.\\Roberto Garello
 is with the Dipartimento di Elettronica, Politecnico di Torino, Italy, email: garello@polito.it.}
\footnote{This research was supported by NSF grant no. CCF-0514881
and grants from Sony, Qualcomm, and the Lee Center for Advanced
Networking.} } \maketitle
\begin{abstract}
Product codes are widely used in data-storage, optical and
wireless applications. Their analytical performance evaluation
usually relies on the truncated union bound,  which provides a low
error rate approximation based on the minimum distance term only.
 In fact, the complete weight enumerator of most product codes remains
unknown. In this paper, concatenated representations are
introduced and applied to compute the complete average enumerators
of arbitrary product codes over a field $F_q$. The split weight
enumerators of some important constituent codes (Hamming,
Reed-Solomon) are studied and used in the analysis. The average
binary weight enumerators of Reed Solomon product codes are also
derived.
Numerical results showing the enumerator behavior are presented.
By using the complete enumerators,  Poltyrev bounds on the maximum
likelihood performance, holding at both high and low error rates,
are finally shown and compared against truncated union bounds and
simulation results.
\end{abstract}

\maketitle

\section{\bf Introduction}
Product codes were introduced by Elias \cite{Elias54} in 1954, who
also proposed to decode them by iteratively (hard) decoding the
component codes. With the invention of turbo codes
\cite{Berrou96}, soft iterative decoding techniques received wide
attention \cite{HagOP96}: low complexity  algorithms for turbo
decoding of product codes were first introduced by Pyndiah in
\cite{Pynd5}. Other efficient algorithms
were recently proposed in \cite{markarian01}
and in \cite{ArgonMclaugh04}.

For product codes, an interesting issue for both theory and
applications regards the analytical estimation of their maximum
likelihood performance. Among other, this analytical approach
allows to: (i) forecast the code performance without resorting to
simulation;
(ii) provide a
benchmark  for testing sub-optimal iterative decoding algorithms;
(iii) establish the goodness of the code, determined by the
distance from theoretical limits.


The analytical performance evaluation  of a maximum-likelihood
decoder requires the knowledge of the code weight enumerator.
Unfortunately, the complete enumerator is unknown for most
families of product codes.
 In these years, some progress has been made in determining the first terms
of product code weight enumerators. The multiplicity
of low weight codewords for an arbitrary linear product code were
computed by Tolhuizen \cite{Tolh02}.
(In our paper, these results will be extended to find the exact
input output weight enumerators of  low weight codewords.)

Even if the first terms can be individuated, the exact
determination of the complete weight enumerator is very hard  for
arbitrary product codes \cite{Tolh02}, \cite{Elk}. By
approximating the number of the remaining codewords by that of a
normalized random code, upper bounds on the performance of binary
product codes using the ubiquitous union bound were shown in
\cite{Tolh98}. However, this approximation is not valid for all
product codes.

In this paper, we will consider the representation of a product
codes as a concatenated scheme with interleaver, and we will
derive the average input-output weight enumerator for linear
product codes over a generic field $F_q$.
When combined with the extended Tolhuizen's result, this will
provide
 a complete approximated  enumerator for the product code. We will
show how it closely approximates the exact weight enumerator.

 Previous work in the literature
(see for example \cite{ChiGar04}, and reference therein) focused
on estimating the product code performance at low error rates via
the truncated union bound, using the enumerator low-weight terms
only. By using the complete approximate enumerator, it is possible
to compute the Poltyrev bound \cite{Polt94}, which establish tight
bounds on the maximum likelihood performance at both high and low
error rates.

The outline of the paper is as follows. In section \ref{Pre}, we
introduce the basic notation and definitions. In section
\ref{sIII}, we extend Tolhuizen results and derive the exact
input-output weight enumerator for product code low-weight
codewords. Product code representation as serial and parallel
concatenated codes with interleavers are introduced in
section~\ref{sIVa}. Uniform interleavers on finite fields with
arbitrary size are discussed in section \ref{sV}. The average
weight enumerators of product codes are then computed in sections
\ref{sIVc}.
The merge with exact low-weight terms, and the discussion of the
combined enumerator properties are performed in
section~\ref{sMerge}.

The computation of  product code average enumerators relies on the
knowledge of the input-redundancy weight enumerators of the
component codes. For this reason, we derive  in section \ref{sVII}
closed form formulas for the  enumerator functions of some linear
codes commonly used in the construction of product codes: Hamming,
extended Hamming, and Reed Solomon codes. We proceed in section
\ref{sVI} to derive the average binary weight enumerators of Reed
Solomon product codes defined on finite fields of characteristic
two.

To support our theory, we present some numerical results. Complete
average enumerators are depicted and discussed in section
\ref{sIXa}. Analytical bounds on the maximum likelihood
performance are shown at both high and low error rates, and
compared against simulation results in section \ref{sIXb}.
Finally, we conclude the paper in section \ref{sX}.

\section{\label{Pre} \bf Preliminaries}

Let $F_q$ be a finite field of $q$ elements, and ${\mathcal C}$ a
$(n_c, k_c, d_c)$ linear code over $F_q$ with codeword length
$n_c$, information vector length $k_c$ and minimum Hamming
distance $d_c$.

The {\bf weight enumerator} ({\bf WE}) of ${\mathcal C}$,
$E_{{\mathcal C}}(h)$, (also called \emph{multiplicity}) is the
number of codewords with Hamming weight $h$:

\eqn{E_{{\mathcal C}}(h)=|\{\mbit{c} \in {{\mathcal C}}: { \tt w}
(\mbit{c})=h \}|,}

\noindent where ${\tt w}(\cdot)$ denotes the symbol Hamming
weight.

For a systematic code ${\mathcal C}$, the {\bf input-redundancy
weight enumerator} ({\bf IRWE}), $R_{{\mathcal C}}(w,p)$, is the
number of codewords with information vector weight $w$, whose
redundancy has weight $p$:

\eqn{R_{{\mathcal C}}(w,p)=|\{\mbit{c=(i|p)} \in {{\mathcal C}}:
{\tt w}(\mbit{i})=w   \;\;\; {\tt w}(\mbit{p})=p \}|,}

If ${\mathcal{T}}=(n_1,n_2)$ is a partition of the $n$ coordinates
of the code into two sets  of size $n_1$ and $n_2$, the
\emph{split weight enumerator} $A^{\mathcal{T}}(w_1,w_2)$ is
number of codewords with Hamming weights $w_1$ and $w_2$ in the
first and second partition, respectively. If $\mathcal{T}$ is an
$(k, n-k)$ partition such the first set of cardinality $k$
constitutes of the information symbol coordinates,
$R(w_1,w_2)=A^{\mathcal{T}}(w_1,w_2)$.

The {\bf input-output weight enumerator} ({\bf IOWE}),
$O_{{\mathcal C}}(w,h)$, is the number of codewords whose Hamming
weight is $h$,  while their information vector has  Hamming weight
$w$:

\eqn{O_{{\mathcal C}}(w,h)=|\{\mbit{c} \in {{\mathcal C}}: {\tt
w}(\mbit{i})=w   \;\;\; {\tt w}(\mbit{c})=h \}|,}

\noindent  For a systematic code,
\begin{equation}
\label{eq:uno} O_{{\mathcal C}}(w,h)=R_{{\mathcal C}}(w,h-w) \;\;.
\end{equation} It is also straight forward that
\begin{equation}
\label{eq:due} E_{{\mathcal C}}(h)=\sum_{w=0}^{k_c} O_{{\mathcal
C}}(w,h) \;\;.
\end{equation}

The \emph{WE function} of ${\mathcal C}$ is defined by this
polynomial in invariant $Y$:
\[{{\mathbb E}}_{{\mathcal C}}(Y)=\sum_{h=0}^{n_c} E_{{\mathcal C}}(h) Y^h \]
while the \emph{IRWE function} and the \emph{IOWE function} of
${\mathcal C}$ are defined by these bivariate polynomials in
invariants $X$ and $Y$:

\eqar{{{\mathbb R}}_{{\mathcal C}}(X,Y)&=&\sum_{w=0}^{k_c}
\sum_{p=0}^{n_c-k_c} R_{{\mathcal C}}(w,p) X^w Y^p,\\
{\mathbb O}_{{\mathcal C}}(X,Y)&=&\sum_{w=0}^{k_c}
\sum_{h=0}^{n_c} O_{{\mathcal C}}(w,h) X^w Y^h. }

\noindent
 These functions  are related by
\begin{equation}
\label{eq:four} {\mathbb O}_{{\mathcal C}}(X,Y)={\mathbb
R}_{{\mathcal C}}\left(XY,Y\right) \end{equation} \noindent and
\begin{equation}
\label{eq:five} {\mathbb E}_{{\mathcal C}}(Y)= {\mathbb
R}_{{\mathcal C}}(Y,Y)={\mathbb O}_{{\mathcal C}}(1,Y).
\end{equation}

In the following, we will denote the coefficient of $X^w Y^h$ in a
bivariate polynomial $f(X,Y)$ by the coefficient function
$\Lambda(f(X,Y), X^w Y^h)$. For example, $O_{{\mathcal
C}}(w,h)=\Lambda({\mathbb O}_{{\mathcal C}}(X,Y),X^w Y^h).$
Similarly, $\Lambda\left({\mathbb O}(X,Y),Y^w\right)$ is the
coefficient of $Y^w$
 in the bivariate polynomial ${\mathbb O}(X,Y)$ and is a univariate polynomial in $X$.

Let the code ${\mathcal C}$  be transmitted by a 2-PSK
constellation over an Additive White Gaussian Noise (AWGN) channel
with a signal-to-noise ratio (SNR) $\gamma$. The symbols
$\Phi_c(\gamma)$ and $\Phi_b(\gamma)$ will
 denote the corresponding codeword error probability
(CEP) and bit error probability (BEP) of a maximum likelihood (ML)
decoder, respectively. These ML performance can be estimated by
computing analytical bounds based on the code enumerators.

The truncated union bound, taking into account the minimum
distance term only, provides  a heuristic approximation commonly
used at high SNR/low CEP:


\begin{equation}
\label{eq:LB} \Phi_c(\gamma) \simeq \frac{1}{2} \,\, E_{{\mathcal
C}}(d_c) \,\, {\tt erfc} \;\; \sqrt{\frac{k_c}{n_c}d_c \gamma}
\,\,\,.
\end{equation}

This formula provides a simple way for predicting the code
performance at very high SNR, where
maximum likelihood error events are mostly due to received noisy
vectors lying in the decoding regions of codewords nearest to the
transmitted one. Anyway, it is not useful in predicting the
performance at low SNR.

Tight bounds on the maximum likelihood codeword error probability
of binary linear codes for AWGN  and binary symmetric channel
(BSC), holding at both low and high SNR, were derived by Poltyrev
in \cite{Polt94}. Other bounds such as the Divsalar simple bound
and the variations on the Gallager bounds are also tight for AWGN
and fading channels \cite{Divs,SasSD03}. These bounds usually
require knowledge of the complete weight enumerator
$E_{\mathcal{C}}(h)$. In this paper, we will apply the Poltyrev
bound by using a complete approximate weight enumerator of the
considered product codes.

Given the codeword error probability, the computation of the bit
error probability may pose a number of technical problems. Let
$\Phi_c(E_{\mathcal{C}}(h),\gamma)$ denote the CEP over a channel
with an SNR $\gamma$ computed by using the weight enumerator
$E_{\mathcal{C}}(h)$.
The bit error probability $\Phi_b(\gamma)$ is derived from the CEP
by computing $\Phi_b(\gamma) = \Phi_c(I_{\mathcal{C}}(h),\gamma)$,
where
%
%
$ I_{\mathcal{C}}(h) = \sum_{w=1}^{k_c} \frac{w}{k_c} O(w,h)$.

A common approximation in the literature is $I_{\mathcal{C}}(h)
\approx \frac{h}{n_c} E_{\mathcal{C}}(h)$. This approximation is
useful if the IOWE $O(\cdot,\cdot)$ is not known but the weight
enumerator WE $E(\cdot)$ is.
Some codes satisfy this approximation with equality: they are said
to possess the \emph{multiplicity property}. This is the case, for
example, of all codes with transitive automorphism groups
(including Hamming and extended Hamming codes) \cite{ChiGar04} or
all maximum distance separable codes (including Reed-Solomon
codes) \cite{ELKMcPWE}.


Let $\mathcal{R}$ and ${\mathcal C}$ be  $(n_r, k_r, d_r)$ and
$(n_c, k_c, d_c)$ linear codes over $F_q$, respectively. The
product code whose component codes are $\mathcal{R}$ and
${\mathcal C}$, $\mathcal{P} \Def \mathcal{R} \times {\mathcal
C}$, consists of all matrices such that each row is a codeword in
$\mathcal{R}$ and each column is a codeword in ${\mathcal C}$.
$\mathcal{P}$ is an $(n_p,k_p,d_p)$ linear code, with parameters
\[ n_p = n_r n_c \;\;\;\;\;\;\;\;\; k_p=k_r k_c \;\;\;\;\;\;\;\; d_p=d_r d_c \]


\section{\label{sIII} \bf Exact IOWE of Product Codes for Low Weight Codewords}

In \cite{Tolh02},
 Tolhuizen showed that in a linear product code $\mathcal{P} =
\mathcal{R} \times {\mathcal C}$ the number of codewords  with
symbol Hamming weight $1 \leq h < h_o$ is:

 \eq{\label{eq:lowenum1}
E_{\mathcal{P}}(h)=\frac{1}{q-1} \sum_{i|h} E_{{\mathcal
C}}(i)E_{\mathcal{R}}(h/i),}
 where, given
\eq{w(d_r,d_c)\Def d_r d_c + \max(d_r \lceil \frac{d_c}{q} \rceil,
d_c \lceil \frac{d_r}{q} \rceil) \nonumber,}

\noindent the weight $h_o$ is \\
\eq{\label{eq:io} h_o=\left\{
                      \begin{array}{ll}
                       w(d_r,d_c)+1, & \hbox{if $q=2$ and both $d_r$ and $d_c$ are odd} \\
                       w(d_r,d_c), & \hbox{otherwise}
                      \end{array}
                    \right.}

In particular, the minimum distance multiplicity of a product code
is given
by 
\eq{\label{newdist}E_{\mathcal{P}}(d_p)=\frac{E_{\mathcal{R}}(d_r)E_{{\mathcal
C}}(d_c)} {q-1}.}


These results are based on the properties of \emph{obvious} (or
\emph{rank-one}) codewords of $\mathcal{P}$, i.e., direct product
of a row and a column codeword \cite{Tolh02}.
Let $\mbit{r} \in \mathcal{R}$ and $\mbit{c} \in {\mathcal C}$,
then an obvious codeword, $\mbit{p} \in \mathcal{P}$, is defined
as \eq{\label{obv}\mbit{p}_{i,j}=\mbit{r}_i \mbit{c}_j,} where
$\mbit{r}_i$ is the symbol in the $i$-th coordinate of $\mbit{r}$
and $\mbit{c}_j$ is the symbol in the $j$-th coordinate of
$\mbit{c}$. It follows that the rank of the $n_c \times n_r$
matrix defined by $\mbit{p}$ is one and the Hamming weight of
$\mbit{p}$ is clearly the product of the Hamming weights of the
component codewords, i.e., \eq{\label{Wprod} {\tt w}(\mbit{p})=
{\tt w}(\mbit{r}) {\tt w}(\mbit{c}).}

 Tolhuizen showed that any codewod with weight smaller
 than $w(d_r,d_c)$ is obvious (Theorem~1, \cite{Tolh02})  (smaller or equal if
 $q=2$ and both $d_r$ and $d_c$ are odd (Theorem~2,
\cite{Tolh02})). The term $\frac{1}{q-1}$ in (\ref{eq:lowenum1})
and (\ref{newdist}) is due to the fact $(\lambda \mbit{r}_i)(
\mbit{c}_j/\lambda)$ are equal for all nonzero $\lambda \in F_q$.

A generalization of Tolhuizen's result to input output weight
enumerators is given in the following theorem.

 \begin{theorem}
 \label{th:IOexth}
Let $\mathcal{R}$ and ${\mathcal C}$ be  $(n_r, k_r, d_r)$ and
$(n_c, k_c, d_c)$ linear codes over $F_q$, respectively. Given the
product code  $\mathcal{P} = \mathcal{R} \times {\mathcal C}$,
%
%
the exact IOWE for codewords with output Hamming weight $1< h <
h_o$ is given by \eq{\label{IOex}O_{\mathcal{P}}(w,h) =
\frac{1}{q-1}\sum_{i|w}\sum_{j|h} O_{\mathcal{R}}(i,j)O_{{\mathcal
C}}(w/i,h/j),} where the sum extends over all factors $i$ and $j$
of $w$ and $h$ respectively, and $h_o$ is given by (\ref{eq:io}).
%
\end{theorem}

\begin{proof}
Let $\mbit{p} \in \mathcal{P}$ be a rank-one codeword, then there
exists a codeword $\mbit{r} \in \mathcal{R}$ and a codeword
$\mbit{c} \in {\mathcal C}$ such that $\mbit{p}_{i,j}=\mbit{r}_i
\mbit{c}_j$.
 The $k_r k_c$ submatrix of information symbols in
$\mbit{p}$ could be constructed from the information symbols in
$\mbit{c}$ and $\mbit{r}$ by (\ref{obv}) for $1 \leq i \leq k_r$ and $1 \leq j
\leq k_c$. It thus follows that the input weight of $\mbit{p}$ is
the product of the input weights of $\mbit{c}$ and $\mbit{r}$ while its output (total) weight is given
by (\ref{Wprod}). Since all codewords with weights $h < h_o$, are rank-one codewords, the theorem follows.
\end{proof}

These results show that both the weight enumerators and the
input-output weight enumerators of product code low-weight
codewords are  determined by the constituent code low-weight
enumerators. This is not the case for larger weights, where the
enumerators of $\mathcal{P}$ are not completely determined by the
enumerators of $\mathcal{R}$ and ${\mathcal C}$ \cite{Tolh02}.

It is important to note the number of rank-one low-weight
codewords is very small, as shown by the following corollary
regarding Reed Solomon (RS) product codes.

\begin{corollary}
\label{cor:one} Let ${\mathcal C}$ be an $(n,k,d)$ Reed Solomon
 code over $F_q$. The weight
enumerator of the product code $\mathcal{P}={\mathcal C} \times
{\mathcal C}$
 has
the following properties,
\eq{E_{\mathcal{P}}(h)=\left\{
                          \begin{array}{ll}
                          1,& \hbox{$h=0$;}\\
                            (q-1)\left({n \choose d}\right)^2, & \hbox{$h=d^2$;} \\
                            0, & \hbox{$d^2 < h < d(d+1)$.}
                          \end{array}
                        \right.}
\end{corollary}

\begin{proof} Let us apply (\ref{eq:lowenum1}).
From the maximum distance separable (MDS) property of RS codes,
$d=n-k+1$ and $n<q$. It follows that $w(d,d)=d(d+1)$. Also
$E_{{\mathcal C}}(d)=(q-1){n \choose d}$. The first obvious
codeword of nonzero weight has weight $d^2$. The next possible
nonzero obvious weight is $d(d+1)$ which is $w(d,d)$.\end{proof}

\begin{example}
Let us consider the ${{\mathcal C}}(7,5,3)$ RS code. The number of
codewords of minimum weight is $E_{{\mathcal C}}(d)=245$. The
complete IOWE function of ${\mathcal C}$ is equal to (this will be
discussed in more detail in section \ref{sVII}):
 \eqar{\nonumber {\mathbb O}_{{\mathcal C}}(X,Y)&=1 + & 35 X
Y^3 + 140 X^2 Y^3 + 70 X^3 Y^3 + 350 X^2 Y^4 +
    700 X^3 Y^4 + 175 X^4 Y^4 \\ \nonumber &&+  2660 X^3 Y^5+2660 X^4 Y^5 +
    266 X^5 Y^5 + 9170 X^4Y^6 + 3668 X^5 Y^6 +
    12873 X^5 Y^7.}

Let $\mathcal{P}$ be the square product code
$\mathcal{P}={\mathcal C} \times {\mathcal C}$. The minimum
distance of $\mathcal{P}$ is $d_p=9$. By (\ref{eq:lowenum1}), its
 multiplicity is $E_{\mathcal{P}}(d_p)=8575$. By applying
Theorem~\ref{th:IOexth}, the input-output weight enumerator for
codewords in $\mathcal{P}$ with output weight $d_p=9$ is given by
\eq{\label{in9}\Lambda({{\mathbb O}}_{\mathcal{P}}(X,Y), Y^9)= 175 X
+ 1400 X^2 + 700 X^3 + 2800 X^4 + 2800 X^6 + 700 X^9.} By
Corollary~\ref{cor:one} there are no codewords in $\mathcal{P}$ with
either weight $10$ or $11$. No information is available for larger
codeword weights $ 12 \leq w \leq 49$.

\end{example}

The following theorem shows that rank-one codewords of a product
code maintain the multiplicity property.

\begin{theorem}
\label{th:prodA} If the codes ${\mathcal C}$ and $\mathcal{R}$
have the multiplicity property and $\mathcal{P}=\mathcal{R} \times
{\mathcal C}$ is their product code, then the subcode constituting
of the rank-one codewords in $\mathcal{P}$ has this property.
\end{theorem}

\begin{proof}
It follows from Th. \ref{th:IOexth} that, for $h \leq h_o$ \eqar{
\nonumber I_{\mathcal{P}}(h)  &=& \frac{1}{q-1}\sum_{w=1}^{k_r
k_c} \frac{w}{k_r k_c} \sum_{i|w}\sum_{j|h}
O_{\mathcal{R}}(i,j)O_{{\mathcal C}}(w/i,h/j)
\\ \nonumber &=& \frac{1}{q-1} \sum_{j|h} \sum_{i=1}^{k_r}
\frac{i}{k_r} O_{\mathcal{R}}(i,j) \sum_{t=1}^{k_c} \frac{t}{k_c}
O_{{\mathcal C}}(t,h/j) \\ \nonumber & =& \frac{1}{q-1}
\frac{h}{n_r n_c}\sum_{j|h} E_{\mathcal{R}}(j) E_{{\mathcal
C}}(h/j)
\\ \nonumber
& =& \frac{h}{n_p} E_{\mathcal{P}}(h),}  which proves the assertion.
\end{proof}
\section{\label{sIV} \bf Average IOWE of Product Codes}

In the previous section, we have shown how to exactly compute
 the product code  IOWE, for low
weight codewords. For higher codeword weights, it is very hard to
find the exact enumerators for an arbitrary product code over
$F_q$.

In this section, we will relax the problem of finding the exact
enumerators, and we will  focus on the computation of  average
weight enumerators over an ensemble of proper concatenated
schemes. To do this:

\begin{enumerate}
\item We will represent  a product code as a concatenated scheme
with a row-by-column interleaver. Two  representations will be
introduced. The first one is the typical serial interpretation of
a product code, while the second one is a less usual parallel
construction. \item We will replace the  row-by-column
interleavers of the schemes by uniform interleavers \cite{Ben96},
acting as the average of all possible interleavers. To do this, we
will introduce and discuss uniform interleavers for codes over
$F_q$. \item We will compute the average enumerator for these
concatenated schemes, which coincide with the scheme enumerators
if random interleavers were used instead of row-by-column ones.
\end{enumerate}


 A code constructed using a random interleaver is no
longer a rectangular product code. However, as we shall see, the
average weight enumerator gives a very good approximation of the
exact weight enumerator of the product code. This will confirm the
experimental results by Hagenauer \emph{et al.} that the error
performance of linear product codes did not differ much if the
row-column interleaver is replaced with a random interleaver
\cite[Sec. IV B]{HagOP96}. We also confirm that numerically in
section~\ref{sIX}.


\subsection{\bf Representing a Product Code as a Concatenated Code}
\label{sIVa}

Let us first study the representation of a product code as a
concatenated scheme with a row-by-column interleaver.

\noindent {\tt Construction 1}

 Given the code ${\mathcal{R}}(n_c,k_c,d_c)$, the augmented code
${\mathcal{R}}^{k_c}$ is obtained by independently
 appending  $k_c$
codewords of $\mathcal{R}$. The code ${\mathcal{R}}^{k_c}$ has
codeword  length $k_c n_r$ and dimension $k_r k_c$. Moreover, its
IOWE function is given by \eqar{\label{outk} {{\mathbb
O}}_{{\mathcal C}}^k(X,Y) \Def {{\mathbb O}}_{{{\mathcal
C}}^k}(X,Y)&=&\left({{\mathbb O}}_{{\mathcal C}}(X,Y)\right)^k.}

See Fig.~\ref{fig:cos1}. The  encoding process may be viewed as if
we are first generating a codeword of ${\mathcal R}^{k_c}$, with
length $k_c n_r$ symbols.  The symbols of this codeword are read
into an $k_c \times n_r$ matrix by rows and read out column by
column. In other words, the symbols of the augmented codeword are
interleaved by a \emph{row-by-column} interleaver. Each column is
then encoded into a codeword in ${\mathcal C}$. The augmented
columns form a codeword in $\mathcal{P}$ of length $n_r n_c$.

\obs{ \label{ser} An $(n_r n_c, k_r k_c, d_r d_c)$ product code
$\mathcal{P} = {\mathcal R} \times {\mathcal C}$ is the serial
concatenation of an $( k_c n_r,   k_c k_r )$ outer code ${\mathcal
R}^{k_c}$ with an $(n_c n_r, k_c n_r)$ inner code ${\mathcal
C}^{n_r}$ through a row-by-column interleaver $\pi$ with length
$N=k_c n_r$. (Equivalently $\mathcal{P} = {\mathcal R} \times
{\mathcal C}$ is the serial concatenation of an $(k_r n_c,k_r
k_c)$ outer code ${\mathcal C}^{k_r}$ with an $(n_c n_r, k_r n_c)$
inner code ${\mathcal R}^{n_c}$ through a row-by-column
interleaver with length $k_r n_c$ respectively.)}

\noindent {\tt Construction 2}


As an alternative, let the coordinates of a systematic product
code be partitioned into four sets as shown in Fig. \ref{PC}.
We can introduce the following parallel representation. \obs{
\label{par} See Figure~\ref{fig:cos2}. An $(n_r n_c, k_r k_c, d_r
d_c)$ product code can be constructed as follows:
\begin{enumerate}
\item Parallel concatenate the $(n_r k_c, k_r k_c)$ code
${\mathcal R}^{k_c}$,  with the $(n_c k_r, k_c k_r)$ code
${\mathcal C}^{k_r}$ through a row-by-column interleaver $\pi_1$
of length $N_1=k_r k_c$. \item Interleave the  parity symbols
generated by ${\mathcal R}^{k_c}$
with a row-by-column interleaver $\pi_2$ of length $N_2=k_c
(n_r-k_r)$. \item  Serially concatenate these  interleaved parity
symbols with the $(n_c (n_r-k_r), k_c(n_r-k_r))$ code ${\mathcal
C}^{n_r-k_r}$.
\end{enumerate}
%

\subsection{\label{sV}\bf Uniform Interleavers over $F_q$}
 Given the two product code representations just introduced, we
would like to substitute the row-by-column interleavers with
uniform interleavers. In this section, we then investigate the
uniform interleaver properties, when the interleaver
 is a symbol based interleaver and the symbols are in $F_q$.
The concept of  uniform interleaver was introduced in \cite{Ben96}
and \cite{Ben98} for
 binary vectors in order to study turbo codes: it is a probabilistic object acting as the average
of all possible interleavers of the given length.  In the binary
case, the number of possible permutations of a vector of length
$L$ and Hamming weight $w$ is ${L \choose w}$. Let us denote by
$V(L,w)$  the probability that a specific vector is  output by the
interleaver when a vector of length $L$ and input $w$ is randomly
interleaved. In this binary case we have
\eq{\label{unif2}V(L,w)=\frac{1}{{L \choose w}}.}

If $\mbit{v}$ is a vector on $F_q$ of length $L$ and the frequency
of occurrence of the $q$ symbols is given by $l_0, l_1, ...,
l_{q-1}$ respectively, then the number of possible permutations is
given by the multinomial coefficient \cite{VLintWilson01}
\eqn{\frac{L!}{l_0! l_1!...l_{q-1}!}.} However, this requires the
knowledge of the occurrence multiplicity of each of the $q$
symbols in the permuted vector.

We then introduce here the notion of \emph{uniform codeword
selector} (UCS). Let us suppose  a specific vector of symbol
weight $w$ and length $L$ is  output from the interleaver
corresponding to a
 certain interleaver input with the same weight. This vector is
 encoded by an $(N,L)$ code ${\mathcal C}$ following the
 interleaver.

We assume that
 all the codewords of ${\mathcal C}$ with input weight $w$ have equal probability of
being chosen at the encoder's output. The UCS picks one of these
codewords (with input weight $w$) at random.
 Thus the probability that a specific codeword is chosen by the UCS is
 \eq{\label{UCS}V(L,w)=\frac{1}{\sum_h O_{{\mathcal C}}(w,h)}=\frac{1}{(q-1)^w {L \choose w}},}
 where $\sum_h O_{{\mathcal C}}(w,h)$ is the total number of
codewords with input weight $w$. This is equivalent to a uniform
interleaver over $F_q$ which identifies codewords by their Hamming
weights.
 It is noticed that for the binary case, the uniform interleaver (\ref{unif2})
 is equivalent to the UCS (\ref{UCS}).
The UCS has the property of preserving the cardinality of the
resulting concatenated code.

\subsection{\bf Computing the Average Enumerators}
\label{sIVc}

\noindent {\tt Construction 1}

 Given the Construction 1 of Obs. \ref{ser} and
Fig.~\ref{fig:cos1}, let us replace the row-by-column interleaver
$\pi$ of length $N=k_c n_r$  with a uniform interleaver over $F_q$
of the same length.
It is easy to show that the average IOWE function of the product
code ${\mathcal P}$ is given by \eq{\label{IOWE1} \bar{{\mathbb
O}}_{{\mathcal P}}(X,Y)=\sum_{w=0}^{k_c n_r}V({k_c n_r, w})
\Lambda \left({\mathbb O}_{{\mathcal R}}^{k_c}(X,Y),Y^w\right)
 \Lambda\left({\mathbb O}_{{\mathcal C}}^{n_r}(X,Y),X^w\right).}

The average weight enumerator function $ \bar{{\mathbb
E}}_{{\mathcal P}}(Y)$ can be computed from $\bar{{\mathbb
O}}_{{\mathcal P}}(X,Y)$ by applying (\ref{eq:five}).

\noindent {\tt Construction 2}

 Given the Construction 2 of Obs. \ref{par} and
Fig.~\ref{fig:cos2}, let us  replace the two row-by-column
interleavers $\pi_1$ of length $N_1=k_r k_c$  and $\pi_2$ of
length $N_2=k_c (n_r-k_r)$, with two uniform interleaver overs
$F_q$ of length $N_1$ and $N_2$, respectively.

We begin by finding the \emph{partition weight enumerator} (PWE)
of the code ${\mathcal D}$ resulting from the parallel
concatenation of ${\mathcal R}^{k_c}$ with ${\mathcal C}^{k_r}$.
We have: \eqar{\label{p1} \bar{{\mathbb P}}_{{\mathcal D}}(W,X,Y)
=\sum_{w=0}^{k_c k_r} V(k_r k_c,w) \Lambda \left({\mathbb
R}_{{\mathcal R}}^{k_c}(W,X),W^w \right) \Lambda\left( {\mathbb
R}_{{\mathcal C}}^{k_r}(W,Y),W^w\right)W^w,} where
$\bar{P}_{{\mathcal D}}(w,x,y)$ is the number of codewords in the
parallel concatenated code
 with weights $w$, $x$ and $y$ in the partitions constituting of information symbols, checks on rows
 and checks on columns respectively, and is given by
\eq{\bar{{\mathbb P}}_{{\mathcal D}}(W,X,Y)=\sum_{w=0}^{k_c k_r}
\sum_{x=0}^{k_c(n_r-k_r)} \sum_{y=0}^{k_r(n_c-k_c)}
 \bar{P}_{{\mathcal D}}(w,x,y)W^w X^x Y^y.}

(Note that $\bar{{\mathbb R}}_{{\mathcal D}}(W,X)=\bar{{\mathbb
P}}_{{\mathcal D}}(W,X,X)$ gives the average IRWE function of a
punctured product code with the checks on checks deleted.)

The partition weight enumerator function of the product code
${\mathcal P}$ is then
 given by
  \eq{\label{p2} \bar{{\mathbb P}}_{{\mathcal P}}(W,X,Y,Z)=\sum_{x=0}^{k_c(n_r-k_r)}
V(k_c(n_r-k_r),x) \Lambda \left( {\mathbb R}_{{\mathcal
C}}^{n_r-k_r}(X,Z),X^x\right) \Lambda \left(\bar{{\mathbb
P}}_{{\mathcal D}}(W,X,Y), X^x\right) X^x.} The PWE,
$\bar{P}_{{\mathcal P}}(w,x,y,z)$, enumerates the codewords with a
weight profile shown in Fig. \ref{PC} and is given by expanding
the PWE function $\bar{{\mathbb P}}_{{\mathcal P}}(W,X,Y,Z)$ as
follows, \eqar{\label{avPWE} \bar{{\mathbb P}}_{{\mathcal
P}}(W,X,Y,Z)= \sum_{w=0}^{k_c
k_r}\sum_{x=0}^{k_c(n_r-k_r)}\sum_{y=0}^{k_r(n_c-k_c)}
\sum_{z=0}^{(n_r-k_r)(n_c-k_c)}\bar{P}_{{\mathcal P}}(w,x,y,z) W^w
X^x Y^y Z^z.} It follows that the average IRWE function of
${\mathcal P}$ is $\bar{{\mathbb R}}_{{\mathcal
P}}(X,Y)=\bar{{\mathbb P}}_{{\mathcal P}}(X,Y,Y,Y)$. Consequently,
the IOWE function $\bar{{\mathbb O}}_{{\mathcal P}}(X,Y)$ can be
obtained via (\ref{eq:four}) and the  weight enumerator function
$\bar{{\mathbb E}}_{{\mathcal P}}(Y)$ via (\ref{eq:five}). By
using (\ref{UCS}),
the cardinality of the code given by $\bar{E}_{{\mathcal P}}(Y)$
is preserved to be $q^{k_r k_c}$.

\section{\bf Merging Exact and Average Enumerators into Combined Enumerators}
\label{sMerge}

The results in the previous section  are now combined with those
of  section~\ref{sIII} reflecting our knowledge of the exact IOWE
of product codes for low weights. Let $h_o$ be defined as in
(\ref{eq:lowenum1}). We introduce a complete IOWE which is equal
to:
\begin{itemize}
\item the exact IOWE for $h <h_o$; \item the average IOWE for $h
\geq h_o$:
\end{itemize}

\eq{\tilde{{\mathbb O}}_{{\mathcal P}}(X,Y)=\sum_{w=0}^{k_r k_c}
\sum_{h=0}^{n_r n_c}\tilde{O}_{{\mathcal P}}(w,h)X^w Y^h,} such
that
\eq{\label{MIOWE} \tilde{O}_{{\mathcal P}}(w,h)=\left\{%
\begin{array}{ll}
   O_{{\mathcal P}}(w,h), & \hbox{$h < h_o$;} \\
    \bar{O}_{{\mathcal P}}(w,h), & \hbox{$h \geq h_o$} \\
\end{array}%
\right.,} where $O_{{\mathcal P}}(w,h)$ is given by Th.
\ref{th:IOexth}, while $\bar{O}_{{\mathcal
P}}(w,h)=\Lambda(\bar{{\mathbb O}}_{{\mathcal P}}(X,Y),X^wY^h)$ is
derived as  in section~\ref{sIVc}.
We will call $\tilde{{\mathbb O}}_{{\mathcal P}}(X,Y)$  the
\textbf{combined input output weight enumerator} (\textbf{CIOWE})
of $\mathcal P$.
The corresponding \emph{combined weight enumerator function}
$\tilde{{\mathbb E}}_{{\mathcal P}}(Y)$  can be computed by
(\ref{eq:five}).


Let us now discuss some properties of the CIOWE. Let $W({\mathcal
C})\Def\{h: E_{{\mathcal C}}(h)\neq 0\}$ be the set of weights
$h$, such that there exists at least one codeword $\mbit{c} \in
{\mathcal C}$ with weight $h$. Observe that the weight of a
product codeword $\mbit{p} \in {\mathcal P}$ is simultaneously
equal to the sum of the row weights and to the sum of the column
weights. We define an integer $h$ a \emph{plausible} weight of
$\mbit{p} \in {\mathcal P}$,
 if $h$ could be simultaneously partitioned into $n_c$ integers
restricted to $W({\mathcal R})$ and into $n_r$ integers restricted
to $W({\mathcal C})$.

 Note however, that not all plausible weights are necessarily in $W({\mathcal P})$.

\begin{theorem}
\label{th:plausible}Suppose ${\mathcal P}={\mathcal R} \times
{\mathcal R}$, (the row code ${\mathcal R}$ is the same as the
column code ${\mathcal C}$), then the set of weights with a
non-zero coefficient in the average weight enumerator of
${\mathcal P}$ derived by either (\ref{IOWE1}) or (\ref{p2}) are
plausible weights for the product code.
\end{theorem}

\begin{proof}
The set of plausible weights of a product code is the set of
weights $h$ such the coefficients of $Y^h$ in both $\left(
{\mathbb E}_{{\mathcal C}}(Y) \right)^{n_r}$ and $\left({\mathbb
E}_{{\mathcal R}}(Y) \right)^{n_c}$ is non-zero. When ${\mathcal
R}={\mathcal C}$, it suffices to show that for any weight $h$ if
the coefficient of $Y^h$ in $\bar{{\mathbb E}}_{{\mathcal P}}(Y)$
is non-zero, then it is also non-zero in $\left( {\mathbb
E}_{{\mathcal C}}(Y) \right)^{n_r}$.

For Construction 1,  let $\bar{{\mathbb E}}_{{\mathcal P}}(Y)$ be
the average weight enumerator derived from (\ref{IOWE1}) by
$\bar{{\mathbb E}}_{{\mathcal P}}(Y)=\bar{{\mathbb O}}_{{\mathcal
P}}(1,Y)$. Since all output weights that appear in $\bar{{\mathbb
O}}_{{\mathcal P}}(1,Y)$ are obtained from $\Lambda\left({\mathbb
O}_{{\mathcal C}}^{n_r}(X,Y),X^w\right)$ then, by (\ref{outk}),
they have nonzero coefficients in $\left( {\mathbb E}_{{\mathcal
C}}(Y) \right)^{n_r}$ and  we are done.

For Construction 2,  let $\bar{{\mathbb E}}_{{\mathcal P}}(Y)$ be
the average weight enumerator derived from (\ref{p2}) by
$\bar{{\mathbb E}}_{{\mathcal P}}(Y)=\bar{{\mathbb P}}_{{\mathcal
P}}(Y,Y,Y,Y)$. Let $\Upsilon(W,Y)=\Lambda \left(\bar{{\mathbb
P}}_{{\mathcal D}}(W,X,Y), X^x\right)$. From (\ref{p1}), it
follows that any exponent with a nonzero coefficients in
$\Upsilon(Y,Y)$ also has a non-zero coefficient in ${\mathbb
R}_{{\mathcal C}}^{k_r}(Y,Y)$ or equivalently ${\mathbb
E}_{{\mathcal C}}^{k_r}(Y)$. Similarly if $\Upsilon'(X,Z)=\Lambda
\left({\mathbb R}_{{\mathcal C}}^{n_r-k_r}(X,Z),X^x\right)X^x$,
then any exponent with a non-zero coefficient in $\Upsilon'(Y,Y)$
also has a non-zero exponent in ${\mathbb E}_{{\mathcal C}}^{n_r -
k_r}$. It follows from (\ref{p2}) that any exponent with a
non-zero coefficient in $\bar{{\mathbb E}}_{{\mathcal P}}(Y)$ also
has a non-zero coefficient in ${\mathbb E}_{{\mathcal C}}^{n_r -
k_r}{\mathbb E}_{{\mathcal C}}^{k_r}$ and we are done.
\end{proof}
In \cite{Tolh98}, the authors approximated the weight enumerator
of the product code by a binomial distribution for \emph{all}
weights greater than $h_o$. Our approach has the advantage that
only \emph{plausible} weights appear in the combined enumerators
of the product code.

\section{\label{sVII} \bf Split weight enumerators}
As seen in the previous section, deriving the CIOWE of the product
code requires the knowledge of the IRWE of the component codes.
In this section we discuss the  weight enumerators   of some codes
which are typically used for product code schemes. In particular,
we show closed form formulas for the IRWE of  Hamming, extended
Hamming, Reed Solomon codes. To do this, it is sometimes easier to
work with  the split weight enumerator (SWE, see definition in
section~\ref{Pre}) of the dual code. The connection between the
IRWE of a code and its dual was established in \cite{weiss}. The
following theorem gives a simplified McWilliams identity relating
the SWE of a linear code with that of its dual code in terms of
Krawtchouk polynomials.

\begin{theorem}
 \label{th:MacW}Let ${\mathcal C}$ be an $(n,k)$ linear code over
$F_q$ and ${\mathcal C}^{\bot}$ be its dual code. Let
$A(\alpha,\beta)$ and $A^{\bot}(\alpha,\beta)$ be the SWEs of
${\mathcal C}$ and ${\mathcal C}^{\bot}$ respectively for an
$(n_1,n_2)$ partition of their coordinates, then
\eqn{A^{\bot}(\alpha,\beta)= \frac{1}{|{\mathcal C}|}
\sum_{w=0}^{n_1}\sum_{v=0}^{n_2}A(w,v) {\mathcal
K}_{\alpha}(w,n_1) {\mathcal K}_{\beta}(v,n_2),} such that for
$\beta=0,1,...,\gamma$, ${\mathcal K}_{\beta}(v,\gamma)=
\sum_{j=0}^{\beta} {{\gamma -v} \choose {\beta -j}} {v \choose j}
(-1)^j (q-1)^{\beta -j}$ is the Krawtchouk polynomial.
\end{theorem}

\begin{proof}
 By a straight forward manipulation
 of the Macwilliams identity for the split weight enumerator
\cite[Ch. 5, Eq. 52]{McSln}\cite{Hwang81}, it follows that for
linear codes and $r=q-1$, \eqar{\nonumber {\mathbb
A}^{\bot}(X,Y)&=&\frac{1}{|{\mathcal C}|} (1+ r X)^{n_1}(1 + r
Y)^{n_2}{\mathbb A} \left( \frac{1-X}{1+r X}, \frac{1-Y}{1+ r Y}
\right)\\ \label{rhseq} &=& \frac{1}{|{\mathcal C}|}
\sum_{w=0}^{n_1} \sum_{v=0}^{n_2} A(w,v)(1- r X)^{n_1-w} (1-X)^{w}
(1- r Y)^{n_2-v} (1-Y)^{v},} where ${\mathbb A}(X,Y)$ and
${\mathbb A}^{\bot}(X,Y)$ are the SWE functions of ${\mathcal C}$
and ${\mathcal C}^{\bot}$ respectively.
 Observing that for a positive integer
$\gamma$ and $0 \leq \beta \leq \gamma$, $(1- r Y)^{\gamma-v}
(1-Y)^{v}= \sum_{\beta=0}^{\gamma} {\mathcal K}_{\beta}(v,\gamma)
Y^{\beta}$ is the generating function for the Krawtchouk
polynomial \cite[Ch. 5, Eq. 53]{McSln} and that $A^{\bot}(\alpha
,\beta)$
 is the
coefficient of $X^{\alpha} Y^{\beta}$ in the right-hand side of (\ref{rhseq}) the result
follows.
\end{proof}

By observing that the roles of the input and the redundancy are
interchanged in the code and its dual, we have:

\begin{corollary}
 \label{cor:corMc} The IRWEs of ${\mathcal C}$ and ${\mathcal
C}^{\bot}$ are related by \eqn{R^{\bot}(\alpha,\beta)=
\frac{1}{|{\mathcal C}|} \sum_{v=0}^{n-k}\sum_{w=0}^{k}R(w,v)
{\mathcal K}_{\beta}(w,k) {\mathcal K}_{\alpha}(v,n-k).}
\end{corollary}

\subsection{Hamming Codes}

The IRWE function of systematic Hamming codes could be derived by observing
that they are the dual code of simplex codes \cite{McSln,LKY04}. 
A recursive equation for evaluating the IRWE of Hamming codes was
given in \cite{SS00}. The following theorem gives a closed form
formula for the IRWE function of  Hamming codes in terms of
Krawtchouk polynomials.

\begin{theorem}
\label{HamIRWE} The IRWE of $(2^m-1, 2^m-m-1,3)$ (systematic)
Hamming codes is \eq{
R_H(\alpha,\beta)=\frac{1}{2^m}\left(\sum_{w=1}^m {m \choose w}
{\mathcal K}_{\beta}(w,m) {\mathcal K}_{\alpha}(2^{m-1}-w,2^m-m-1)
\right.+ \left. {m \choose \beta} {2^m-m-1 \choose
\alpha}\right)\nonumber.}
\end{theorem}
\begin{proof}
By observing that the IRWEF of the $(2^m-1, m,2^{m-1})$ simplex
code is ${\mathbb R}_s(X,Y)=1+\sum_{w=1}^{m} {m \choose w} X^w
Y^{2^{m-1}-w}$. Using Cor. \ref{cor:corMc} and observing that
${\mathcal K}_{\beta}(0,m)= {m \choose \beta}$, we obtain the
result.
\end{proof}

\subsection{Extended Hamming Codes}

Extended Hamming codes were studied in \cite{ChiGar04}, where it
was shown they possess the multiplicity property, and closed-form
formulas for their input-output multiplicity  were provided. The
following theorem shows a closed expression for their IRWE
function in terms of Krawtchouk polynomials.

\begin{theorem}
\label{th:EHamIRWE}A closed form formula for the IRWE of the
$(2^m, 2^m-m-1,4)$ Extended Hamming codes is \eqar{\nonumber
R_{EH}(\alpha,\beta)=\frac{1}{2^{m+1}} \left(\sum_{w=1}^m {{m+1}
\choose w} {\mathcal K}_{\beta}(w,m+1)\right. {\mathcal
K}_{\alpha}(2^{m-1}-w,2^m-m-1)\\ \nonumber + \left.{{m+1} \choose
\beta} {2^m-m-1 \choose \alpha}
\left(1+(-1)^{\alpha+\beta}\right)\right).}
\end{theorem}

\begin{proof}
By observing that the extended Hamming codes are the duals of the
$(2^m,m+1,2^{m-1})$ first order Reed Muller (RM) codes whose IRWE
function could be shown to be ${\mathbb
R}(X,Y)=1+X^{m+1}Y^{2^m-m-1}+\sum_{\alpha=1}^m {{m+1} \choose
\alpha} X^{\alpha} Y^{2^{m-1}-\alpha}.$ By Cor. \ref{cor:corMc}
and observing that ${\mathcal K}_{\beta}(\gamma,\gamma)={\gamma
\choose \beta} (-1)^ {\beta}$ the result follows.
 \end{proof}

Note that the WE of extended Hamming (EH) codes could also be
derived from that of Hamming (H) codes
 by using the well known relation \cite{McSln}, $E_{EH}(h)=E_{H}(h)+
E_{H}(h-1)$ if $h$ is even and is zero otherwise.
It follows that
\eq{R_{EH}(\alpha,\beta)=\left\{
                      \begin{array}{ll}
                       R_{H}(\alpha,\beta)+R_{H}(\alpha,\beta-1), & \hbox{$\alpha+\beta$ is even;} \\
                        0, & \hbox{otherwise}
                      \end{array}
                    \right.}

\subsection{Reed Solomon Codes}
Reed Solomon codes are maximum distance separable (MDS) codes
\cite{McSln}. The SWE of MDS codes was recently studied by
El-Khamy and McEliece in \cite{ELKMcPWE}, where this theorem was
proved:

\begin{theorem}
\cite{ELKMcPWE}
 The SWE of MDS codes is given by \label{lem:split}
\eqn{A^{\mathcal{T}}(w_1,w_2) = E(w_1+w_2) \frac{ {n_1 \choose
w_1} {n_2 \choose w_2}}{{n \choose {w_1+w_2}}}.}
\end{theorem}

 It follows that the IRWE of an $(n,k)$ systematic RS
code is given by:

\[R_{RS}(\alpha,\beta)=E(\alpha+\beta) \frac{{k \choose \alpha}{{n-k} \choose {\beta}}}{{n \choose
\alpha+\beta}}.\]

\section{\label{sVI} \bf Average Binary IRWE of product Reed Solomon Codes}

 Recently, new techniques for decoding Reed Solomon codes  beyond half the minimum
distance were derived in \cite{GS}, and algebraic soft decision
algorithms were proposed (see \cite{ElkMcDimacs} and references
therein). In this section we derive a number of results on RS
product codes and their binary image.

 Let $\mbit{d}(X)=\sum_{i=0}^{k-1} d_i X^i$ be a data polynomial
over $F_q$. Then an $(n,k)$ Reed Solomon code is generated by
evaluating the data polynomial $\mbit{d}(X)$ at $n$ distinct
elements of the field forming a set called the \emph{support} set of
the code $S=\{ \alpha_0, \alpha_1,..., \alpha_{n-1}\}$. The
generated codeword is $\mbit{c}=\{\mbit{d}(\alpha_0),
\mbit{d}(\alpha_1),..., \mbit{d}(\alpha_{n-1})\}$. The resulting
code is not systematic. In the following theorem we show how a
product of two RS codes can be generated by polynomial evaluation of
a bivariate polynomial.

\begin{theorem}
\label{th:RS} Let the $k_r k_c$ data symbols $d_{i,j}$
 be given by the bivariate polynomial $D(X,Y) = \sum_{i=0}^{k_r-1} \sum_{j=0}^{k_c-1} d_{i,j} X^i Y^j$. Let the
 support set of the row and column codes, ${\mathcal R}$ and ${\mathcal C}$ respectively,
  be given by $S_r=\{\alpha_0, \alpha_1,..., \alpha_{n_r-1}\}$
and $S_c=\{\beta_0, \beta_1, ..., \beta_{n_c-1}\}$ respectively.
Then the product code ${\mathcal P}={\mathcal R} \times {\mathcal
C}$ is generated by ${\mathcal P}_{i,j}= D(\alpha_i,\beta_j)$ for
$i=0,..,n_r-1$ and $j=0,...,n_c-1$.
\end{theorem}

\begin{proof}
To prove that ${\mathcal P}$ is really the product of ${\mathcal
R}$ and ${\mathcal C}$, we prove that each row, $r$, ${\mathcal
P}_{*,r}=\{{\mathcal P}_{i,j}: i=\{0,...,n_r-1\} \& j=r\}$ is a
codeword in ${\mathcal R}$ and each column $c$, ${\mathcal
P}_{c,*}=\{{\mathcal P}_{i,j}: j=\{0,...,n_c-1\} \& i=c\}$ is a
codeword in ${\mathcal C}$. The $r$-th row is given by ${\mathcal
P}_{r,*}=\{D(\alpha_0,\beta_r), D(\alpha_1,\beta_r),...,
D(\alpha_{n_r-1},\beta_r)\}$. Observe that $D(\alpha_i,\beta_r) =
\sum_{v=0}^{k_r-1} \sum_{w=0}^{k_c-1} d_{v,w} (\alpha_i)^v
(\beta_r)^w = \sum_{v=0}^{k_r-1} \left(\sum_{w=0}^{k_c-1} d_{v,w}
(\beta_r)^w \right)(\alpha_i)^v.$ Let
$\gamma^r_v=\sum_{w=0}^{k_c-1} d_{v,w} (\beta_r)^w $ and
$\gamma^r=\{\gamma^r_v: v=0,...,k_r-1\}$, then $\gamma^r$ forms
the information vector which is encoded into the $r$-th row
${\mathcal P}_{*,r}=\{D'(\alpha_0),
D'(\alpha_1),...,D'(\alpha_{n_r-1})\}$ by
$D'(X)=\sum_{i=0}^{k_r-1} \gamma^r_i X^i$. This proves that
${\mathcal P}_{*,r}$ is a codeword in ${\mathcal R}$.

Similarly, any column $c$ could be expressed as ${\mathcal P}_{c,*}=\{D''(\beta_0), D''(\beta_1),...,D''(\beta_{n_c-1})\}$, \\
where $D''(X)=\sum_{j=0}^{k_c-1} \delta^c_j X^j$ and
$\delta^c_j=\sum_{i=0}^{k_r-1} d_{i,j} (\alpha_c)^i$ is the $j$-th
element in the information vector for column $c$. Thus each column
is a codeword in ${\mathcal C}$.

Since the cardinality of this code is $q^{k_r k_c}$ we are done.
\end{proof}
(This proof also gives insight how algebraic soft decision
decoding
can be used for iteratively decoding the component RS codes of the
product code.)

Given the product of Reed Solomon codes  defined over a field of
characteristic two, it is often the case that the binary image of
the code is transmitted over a binary-input channel. The
performance would thus depend on the binary weight enumerator of
the component RS codes, which in turn depends on the basis used to
represent the $2^m$-ary symbols as bits. Furthermore, it is very
hard to find the exact binary weight enumerator of a RS code for a
specific basis representation \cite{KasamiLin88}. The average
binary image of a class of generalized RS codes has been studied
in \cite{Retter91}. The average binary image for codes, defined
over finite fields of characteristic two, was derived by assuming
a binomial distribution of the bits in the non-zero symbols in
\cite{ELKMcAller}. Let ${\mathcal C}_b$ denote the binary image of
an $(n,k)$ code ${\mathcal C}$ which is defined over the finite
field $F_{2^m}$. Let ${\mathbb E}_{{\mathcal C}}(Y)$ be the weight
enumerator function of ${\mathcal C}$. Then the average weight
enumerator of the $(n m,k m)$ code ${\mathcal C}_b$ is given by
\cite{ELKMcAller} \eq{\bar{{\mathbb E}}_{{\mathcal
C}_b}(Y)={\mathbb E}_{{\mathcal C}}(\Psi(Y)),} where
$\Psi(Y)=\frac{1}{2^m-1}((1+Y)^m-1)$ is the generating function of
the bit distribution in a non-zero symbol. We assume that the
distribution of the non-zero bits in a non-zero symbol follows a
binomial distribution and that the non-zero symbols are
independent. If the coordinates of the code ${\mathcal C}$ are
split into $p$ partitions, then there is a corresponding
$p$-partition of the coordinates of ${\mathcal C}_b$, where each
partition in ${\mathcal C}_b$ is the binary image of a partition
in ${\mathcal C}$. By independently finding the binary image of
each partition,
 the average partition weight enumerator  of ${\mathcal C}_b$ could be derived as in the following lemma.

\begin{lemma}
\cite{ELKMcPWE}Let ${\mathbb P}_{{\mathcal P}}(W,X,Y,Z)$ be the
PWE function of a code ${\mathcal P}$ defined over $F_{2^m}$. The
average PWE of the binary image ${\mathcal P}_b$ is $\bar{{\mathbb
P}}_{{\mathcal P}_b}(W,X,Y,Z) = {\mathbb P}_{{\mathcal
P}}\left(\Psi(W),\Psi(X),\Psi(Y),\Psi(Z)\right).$
\end{lemma}

\begin{corollary}
\label{cor:bin} If $\tilde{{\mathbb R}}_{{\mathcal P}}(X,Y)$ is
the combined IRWE of the $(n_p,k_p)$ product code ${\mathcal P}$
defined over $F_{2^m}$, then the combined IRWE of its binary image
is \[\tilde{{\mathbb R}}_{{\mathcal P}_b}(X,Y)=\tilde{{\mathbb
R}}_{{\mathcal P}}(\Psi(X),\Psi(Y)),\] \noindent  where
\[\Psi(X)=\frac{1}{2^m-1}((1+X)^m-1)\]
\noindent and \[ \tilde{{\mathbb R}}_{{\mathcal P}_b}(X,Y)=
\sum_{x=0}^{k_p m}\sum_{y=0}^{n_p m-k_p m}R_{{\mathcal P}_b}(x,y)
X^x Y^y.\]
\end{corollary}

Note this same formula does not hold in the case of the IOWE.
However, the binary IOWE could be derived from the binary IRWE by
using (\ref{eq:four}).

\section{\label{sIX} \bf Numerical Results}

In this section we show some numerical results supporting our
theory. The combined input output enumerators of some product
codes are investigated in section~\ref{sIXa}. Analytical bounds to
ML performance are computed and discussed in section~\ref{sIXb}.
Hamming codes, extended Hamming codes and Reed Solomon codes are
considered as constituent codes.

\subsection{\bf Combined Input Output Weight Enumerators}
\label{sIXa}

\begin{example}
Let us consider the extended Hamming code (8,4).
 From Th. \ref{th:EHamIRWE}, its IOWE function is

 \[{\mathbb O}_{EH}(X,Y)=1+4 X Y^4+6 X^2 Y^4+4 X^3 Y^4+X^4
 Y^8\;\;\;.\]

Let us now study the square product code $(8,4)^2$. By applying
(\ref{IOWE1}) we can derive the average weight enumerator function
obtained with the serial concatenated representation. By rounding
to the nearest integer, we obtain:


\eqar{ {\mathbb E}_{{\mathcal P}}(Y)&=1&+3 Y^8+27 Y^{12}+107
Y^{16}+604 Y^{20}+3153 Y^{24}+ 13653 Y^{28}+30442 Y^{32}+ \nonumber \\
&& +13653 Y^{36}+3153 Y^{40}+ 604 Y^{44}+107 Y^{48}+27 Y^{52}+3
Y^{56}+Y^{64}. \nonumber}


By (\ref{p2}), we can derive the average weight enumerator
function obtained with the parallel concatenated representation.
We obtain:

\eqar{ {\mathbb E}_{\mathcal P}(Y)&=1&+2 Y^8+26 Y^{12}+98
Y^{16}+568 Y^{20}+ 3116 Y^{24}+13780 Y^{28}+30353 Y^{32}+
\nonumber
\\&& + 13780 Y^{36}+ 3116 Y^{40}+568 Y^{44}+98 Y^{48}+26 Y^{52}+2
Y^{56}+Y^{64}}

 \noindent (For space limitations we do not show the IOWE functions.) Note that
all codewords are of \emph{plausible} weights as expected from Th.
\ref{th:plausible}.
 It could be checked that in both cases, the cardinality of the code (without rounding) is preserved to be $2^{16}$.
In general, the parallel representation gives more accurate
results than the serial one, and will be used for the remaining
results in this paper.

For low-weight codewords, we can compute the exact IOWE.
 By Th.\ref{th:IOexth}, the exact IOWE of the product code for weights less than $h_o=24$
 is equal to
  \[ {\mathbb O}_{{\mathcal P}}(X,Y)=1+16 X Y^{16}+48 X^2 Y^{16}+32 X^3 Y^{16}+
36 X^4 Y^{16}+48 X^6 Y^{16}+16 X^9 Y^{16}.\]

 It follows that the
combined weight enumerator function  for this product code is

\eqar{\tilde{\mathbb E}_{\mathcal P}(Y)&=1&+196Y^{16}+3116 Y^{24}+
13781 Y^{28}+30353 Y^{32} \nonumber \\ && + 13781 Y^{36}+3116
Y^{40}+568 Y^{44}+98 Y^{48}+ 26 Y^{52}+2 Y^{56}+Y^{64}.\nonumber }

A symmetric weight enumerator of the component codes implies a
symmetric one for the product code. Thus, by the knowledge of the
exact coefficients of exponents less than $24$, $\tilde{{\mathbb
E}}_{{\mathcal P}}(Y)$ could be improved by setting the
coefficients of $Y^{54},Y^{52}$ and $Y^{56}$ to be zero and
adjusting the coefficients of the middle exponents such that the
cardinality of the code is preserved. We obtain:

\eqar{ \tilde{{\mathbb E}'}_{{\mathcal P}}(Y)&=1&+196 Y^{16}+3164
Y^{24}+ 13995 Y^{28}+30824 Y^{32} \nonumber \\&& + 13995
Y^{36}+3164 Y^{40}+196 Y^{48}+Y^{64}, \label{eq:ex1}}

In this case, the exact weight enumerator can be found by
exhaustively generating the 65536 codewords of the product code,
and it is equal to:

\eqar{ {\mathbb E}_{{\mathcal P}}(Y)&=1&+196 Y^{16}+4704 Y^{24}+
10752 Y^{28}+34230 Y^{32} \nonumber \\ && + 10752 Y^{36}+4704
Y^{40}+196 Y^{48}+Y^{64}. \nonumber}

It can be verified that the combined weight enumerator
(\ref{eq:ex1}) gives a very good approximation of this exact
weight enumerator.


\end{example}

\begin{example}
The combined weight enumerator  of the extended Hamming product
code $(16,11)^2$, computed by applying (\ref{p1}) and (\ref{p2}),
is depicted in Fig. \ref{AWE1611EH}. It is observed that for
medium weights, the distribution is close to that of random codes,
which is  given by \eqn{E(w)=q^{-(n_p-k_p)}{n_p \choose
w}(q-1)^w,}
 except that only plausible weights exist.

\end{example}

\begin{example}
The combined symbol weight enumerator of the $(7,5)^2$ product RS
codes over $F_8$, computed by applying (\ref{p1}) and (\ref{p2}),
is shown in Fig. \ref{fRS75}.
 It can be  observed that the weight enumerator approaches
that of a random code over $F_8$ for large weights. The average binary weight enumerator
 of the $(147,75)$ binary image, obtained by applying Corollary \ref{cor:bin}, is shown in Fig.
\ref{fRS752}. It is superior to a random code at low weights and
then, as expected, approaches that of a binary random code.
\end{example}

\subsection{\bf Maximum Likelihood Performance}
\label{sIXb}

In this section, we  investigate  product code performance. The
combined weight enumerators are used to compute the Poltyrev bound
\cite{Polt94}, which gives tight analytical bounds to maximum
likelihood performance at both high and low error rates. For
proper comparison, truncated union bound approximation and
simulation results are also considered.

\begin{example}
 The codeword error rate (CER) and the bit error rate (BER) performance  of two Hamming product codes ($(7,4)^2$ and $(31,26)^2$) are shown in Fig.
 \ref{hamperf}. We  have depicted:
 \begin{itemize}
 \item The Poltyrev bounds on ML performance (P on the plots),
 obtained by using the combined weight enumerator computed via (\ref{p2}).
 \item The truncated union bound (L on the plots), approximating the ML performance at low error rates, and computed  from the
 minimum distance term  via (\ref{eq:LB}).
 \item The simulated performance of iterative decoding (S on the plots), corresponding to 15 iterations of
 the BCJR algorithm on the constituent codes trellises (\cite{HagOP96},\cite{ChiGar04}).
 \end{itemize}

\noindent
 By looking at the results, we can observe that:

\begin{itemize}
\item The combined weight enumerators derived in this paper, in
conjunction with the Poltyrev bound, provide very tight analytical
bounds on the performance of maximum likelihood decoding also at
low SNRs (where the truncated union bound does not provide useful
information). \item For the $(7,4)^2$ code the exact enumerator
can be exhaustively computed, and the exact Poltyrev bound is
shown in the figure. It is essentially identical to the bound
computed with the combined weight enumerator. \item The ML
analytical bounds provide very useful information also for
iterative decoding performance. In fact,  the penalty paid by
iterative decoding with respect to ideal ML decoding is very
limited, as shown in the figure (feedback coefficients for
weighting the extrinsic information and improve iterative decoding
has been employed, as explained in \cite{ChiGar04}).
\end{itemize}
\end{example}



\begin{example}
The performance of the extended Hamming product code $(32,26)^2$
is investigated in Fig.~\ref{fEH32}. Also in this case, the
tightness of the bounds is demonstrated, for both the CER and the
BER. With the aid of the Poltyrev bound for the BSC channel, hard
ML bounds have also been plotted. It is shown that soft ML
decoding on the AWGN channel offers more than $2$ dB coding gain
over hard ML decoding.
\end{example}

\begin{example}
In Fig.~\ref{fEHfH}, the performance of soft and hard ML decoding
of various Hamming and extended Hamming codes are studied and
compared.
 As expected, the EH product codes show  better performance than Hamming product codes of the same
length due to their larger minimum distance and lower rate. (For
the $(7,4)^2$ Hamming product code and the $(8,4)^2$ extended
Hamming product code, it is observed that the bounds using our
combined weight enumerator overlapped with ones using the exact
weight enumerators, which can be calculated exhaustively in these
cases.)
\end{example}

\begin{example}
The  performance of the binary image of some Reed Solomon product
codes,   for both soft and hard decoding,  are investigated in
Fig. \ref{fRS}, where the Poltyrev bound has been plotted. As
expected, soft decoding has about $2$ dB of gain over hard
decoding. It can be observed that these product codes have good
performance at very low error rates (BER lower than $10^{-9}$),
where no error floor appears.

 It is
well known that the sphere packing bound provides a lower bound to
the performance achievable by a code with  given code-rate and
codeword length \cite{ValFos}. The discrete-input further limitation
occurring when using a given PSK modulation format was addressed in
\cite{SPB}. The distance of the code performance from this
theoretical limit can be used an indicator of the code goodness.

Let us consider, for example, the $(15,11)^2$ RS product code,
corresponding to a (900,484) binary code. By looking at the
Poltyrev bound plotted in Fig. \ref{fRS}, this code achieves a
BER=$10^{-10}$ for a signal-to-noise ratio $\gamma \simeq
2.2\;$dB. By computing the PSK sphere packing bound for this
binary code, we obtain a value of about 1.9 dB for BER=$10^{-10}$.
This means that this RS product code is within 0.3 dB from the
theoretical limit, which is a very good result at these low error
rates.

\end{example}

\section{\label{sX}\bf Conclusions}
The average weight enumerators of product codes were studied in
this paper. The problem was relaxed by considering  proper
concatenated
 representations, and assuming random interleavers over $F_q$
instead of  row-by-column interleavers. The exact IOWE for
low-weight codewords were also derived by extending Tolhuizen
results. By combining exact values and average values, a complete
combined weight enumerator was computed. This enables us to study
the ML performance of product codes at both low and high SNRs by
applying the Poltyrev bound. The computation of average
enumerators requires knowledge of the constituent code
enumerators. Closed form formulas for the input redundancy
enumerators of some popular codes
were shown.
The binary weight enumerator of ensemble of product RS codes was
also derived.

 The combined weight enumerators of Hamming and Reed Solomon
product codes were numerically computed and discussed. Using the
combined enumerators,  tight bounds on the ML performance of product
codes over AWGN channels were derived by using the Poltyrev bounds.
The tightness of the bounds were demonstrated by comparing them to
both  truncated union bound approximations and simulation results.

In particular, Reed Solomon product codes show excellent
performance. Reed Solomon codes are widely used in wireless, data
storage, and optical systems due to their burst-error correction
capabilities. The presented techniques allow to analytically
estimated  Reed Solomon product codes performance, and show they
are very promising as Shannon-approaching solutions down to very
low error rates without error floors. This suggests the search for
low-complexity soft decoding algorithms for Reed Solomon codes as
a very important research area in the near future.


\section*{Acknowledgment}
The authors are grateful to Robert J. McEliece for very useful
discussions.

\bibliographystyle{IEEEtran}
\bibliography{ElkGar3.bib}

\begin{thebibliography}{10}
\providecommand{\url}[1]{#1}
\csname url@rmstyle\endcsname
\providecommand{\newblock}{\relax}
\providecommand{\bibinfo}[2]{#2}
\providecommand\BIBentrySTDinterwordspacing{\spaceskip=0pt\relax}
\providecommand\BIBentryALTinterwordstretchfactor{4}
\providecommand\BIBentryALTinterwordspacing{\spaceskip=\fontdimen2\font plus
\BIBentryALTinterwordstretchfactor\fontdimen3\font minus
  \fontdimen4\font\relax}
\providecommand\BIBforeignlanguage[2]{{%
\expandafter\ifx\csname l@#1\endcsname\relax
\typeout{** WARNING: IEEEtran.bst: No hyphenation pattern has been}%
\typeout{** loaded for the language `#1'. Using the pattern for}%
\typeout{** the default language instead.}%
\else
\language=\csname l@#1\endcsname
\fi
#2}}

\bibitem{Elias54}
P.~Elias, ``Error-free coding,'' \emph{IRE Trans. Inform. Theory}, vol. IT-4,
  pp. 29–--37, Sept 1954.

\bibitem{Berrou96}
C.~Berrou and A.~Glavieux, ``Near-optimum correcting coding and decoding: Turbo
  codes,'' \emph{IEEE Trans. Commun.}, vol.~44, pp. 1261--–1271, Oct. 1996.

\bibitem{HagOP96}
J.~Hagenauer, E.~Offer, and L.~Papke, ``Iterative decoding of binary block and
  convolutional codes,'' \emph{{IEEE} Trans. Inform. Theory}, vol.~42, pp.
  429--449, Mar 1996.

\bibitem{Pynd5}
R.~Pyndiah, ``Near optimum decoding of product codes: block turbo codes,''
  \emph{{IEEE} Trans. Commun.}, vol.~46, no.~8, pp. 1003--1010, August 1998.


\bibitem{markarian01}
S.A. Hirst, B.~Honary, and G.~Markarian,
\newblock ``{Fast Chase Algorithm with an Application in Turbo Decoding}'',
\newblock {\em IEEE Trans. Communications},  pp. 1693--1699,
  Oct. 2001.

\bibitem{ArgonMclaugh04}
C.~Argon and S.~W. McLaughlin, ``An efficient chase decoder for turbo product
  codes,'' \emph{{IEEE} Trans. Commun.}, vol.~52, no.~6, pp. 896--898, June
  2004.


\bibitem{Tolh02}
L.~Tolhuizen, ``More results on the weight enumerator of product codes,''
  \emph{{IEEE} Trans. Inform. Theory}, vol.~48, no.~9, pp. 2573–--2577, Sep.
  2002.
  
  \bibitem{Elk}
M.~El-Khamy, ``The average weight enumerator and the maximum likelihood
  performance of product codes,'' in \emph{International Conference on Wireless
  Networks, Communications and Mobile Computing, WirelessCom Information Theory
  Symposium}, June 2005.

\bibitem{Tolh98}
L.~Toluizen, S.~Baggen, and E.~Hekstra-Nowacka, ``Union bounds on the
  performance of product codes,'' in \emph{Proc. of ISIT 1998}, Cambridge, MA,
  USA, 1998.

\bibitem{ChiGar04}
F.~Chiaraluce and R.~Garello, ``Extended hamming product codes analytical
  performance evaluation for low error rate applications,'' \emph{IEEE
  Trans. on Wireless Commun.}, vol.~3, pp. 2353--2361, Nov. 2004.



\bibitem{Polt94}
G.~Poltyrev, ``Bounds on the decoding error probability of binary linear codes
  via their spectra,'' \emph{{IEEE} Trans. Inform. Theory}, vol.~40, no.~4, pp.
  1284--1292, July 1994.

\bibitem{Divs}
D.~Divsalar, ``A simple tight bound on error probability of block codes with
  application to turbo codes,'' TMO Progress Report, NASA,JPL, Tech. Rep.
  42--139, 1999.

\bibitem{SasSD03}
I.~Sason, S.~Shamai, and D.~Divsalar, ``Tight exponential upper bounds on the
  {ML} decoding error probability of block codes over fully interleaved fading
  channels,'' \emph{{IEEE} Trans. Commun.}, vol.~51, no.~8, pp. 1296--1305,
  Aug. 2003.

\bibitem{ELKMcPWE}
M.~El-Khamy and R.~J. McEliece, ``The partition weight enumerator of {MDS}
  codes and its applications.'' in \emph{2005 IEEE International Symposium on
  Information Theory, Adelaide, Australia}, Sept 2005.

\bibitem{Ben96}
S.~Benedetto and G.~Montorsi, ``Unveiling turbo codes: Some results on parallel
  concatenated coding schemes,'' \emph{{IEEE} Trans. Inform. Theory},
  vol.~42, no.~3, pp. 409–--428, Mar. 1996.

\bibitem{Ben98}
S.~Benedetto, D.~Divsalar, G.~Montorsi, and F.~Pollara, ``Serial concatenation
  of interleaved codes: Performance analysis, design and iterative decoding.''
  \emph{{IEEE} Trans. Inform. Theory}, vol.~44, no.~3, pp. 909–--926, May 1998.

\bibitem{VLintWilson01}
J.~H. van Lint and R.~M. Wilson, \emph{A {C}ourse in {C}ombinatorics},
  2nd~ed.\hskip 1em plus 0.5em minus 0.4em\relax Cambridge: Cambridge U. Press,
  2001.
  
  \bibitem{weiss}
C.~Wei$\beta$, C.~Bettstetter, and S.~Riedel,
\newblock ``{Code Construction and Decoding of Parallel Concatenated
  Tail-Biting Codes}'',
\newblock {\em IEEE Trans. Inform. Theory},  pp. 366--386, Jan.
  2001.



\bibitem{McSln}
F.~J. MacWilliams and N.~J. Sloane, \emph{The Theory of Error Correcting
  Codes}.\hskip 1em plus 0.5em minus 0.4em\relax Amsterdam: North Holland,
  1977.

\bibitem{Hwang81}
T.-Y. Hwang, ``A relation between the row weight and column weight
  distributions of a matrix,'' \emph{{IEEE} Trans. Inform. Theory}, 
vol.~27, no.~2, pp. 256 -- 257, Mar. 1981.
  

\bibitem{LKY04}
H.~feng Lu, P.~V. Kumar, and E.~hui Yang, ``On the input-output weight
  enumerators of product accumulate codes,'' \emph{{IEEE} Commun. Lett.},
  vol.~8, no.~8, Aug 2004.

\bibitem{SS00}
I.~Sason and S.~Shamai, ``Bounds on the error probability for block and
  turbo-block codes,'' \emph{Annals of Telecommunications}, vol.~54, no.~3-4,
pp.~183 - 200, March - April 1999. 

  
  \bibitem{GS}
V.~Guruswami and M.~Sudan, ``Improved decoding of {R}eed-{S}olomon codes and
  algebraic geometry codes,'' \emph{{IEEE} Trans. Inform. Theory}, vol.~45,
  no.~6, pp. 1757–--1767, Sept. 1999.

\bibitem{ElkMcDimacs}
M.~El-Khamy and R.~J. McEliece, ``Interpolation multiplicity assignment
  algorithms for algebraic soft-decision decoding of {R}eed-{S}olomon codes,''
  \emph{{AMS}-DIMACS volume on Algebraic Coding Theory and Information Theory},
  vol.~68, 2005.

\bibitem{KasamiLin88}
T.~Kasami and S.~Lin, ``The binary weight distribution of the extended
  $(2^m,2^m-4)$ code of the {R}eed {S}olomon code over {GF}($2^m$) with
  generator polynomial $(x-\alpha)(x-\alpha^2)(x-\alpha^3)$.'' \emph{Linear
  Algebra Appl.}, pp. 291–--307, 1988.

\bibitem{Retter91}
C.~Retter, ``The average binary weight enumerator for a class of generalized
  {R}eed-{S}olomon codes,'' \emph{{IEEE} Trans. Inform. Theory}, vol.~37,
  no.~2, pp. 346--349, March 1991.

\bibitem{ELKMcAller}
M.~El-Khamy and R.~J. McEliece, ``Bounds on the average binary minimum distance
  and the maximum likelihood performance of {R}eed {S}olomon codes,'' in
  \emph{42nd Allerton Conf. on Communication, Control and Computing}, 2004.
  
   \bibitem{ValFos}
  A.~Valembois and M.~Fossorier, ``Sphere-Packing Bounds Revisited for Moderate
Block Lengths,'' \emph{{IEEE} 
  Trans. Inform. Theory}, vol.~50, no.~12, pp. 2998--3014,  Dec. 2004.
  
  \bibitem{SPB}
  G.~Beyer, K.~Engdahl, and KSh.~Zigangirov, ``Asymptotic analysis and comparison 
  of two coded modulation schemes using PSK signaling - Part I,''  \emph{{IEEE} 
  Trans. Inform. Theory}, vol.~47, no.~7, pp. 2782--2792,  Nov. 2001.
  
 
  

\end{thebibliography}


\clearpage

\begin{figure}
\centering 
\includegraphics[clip,scale=0.8,angle=270]{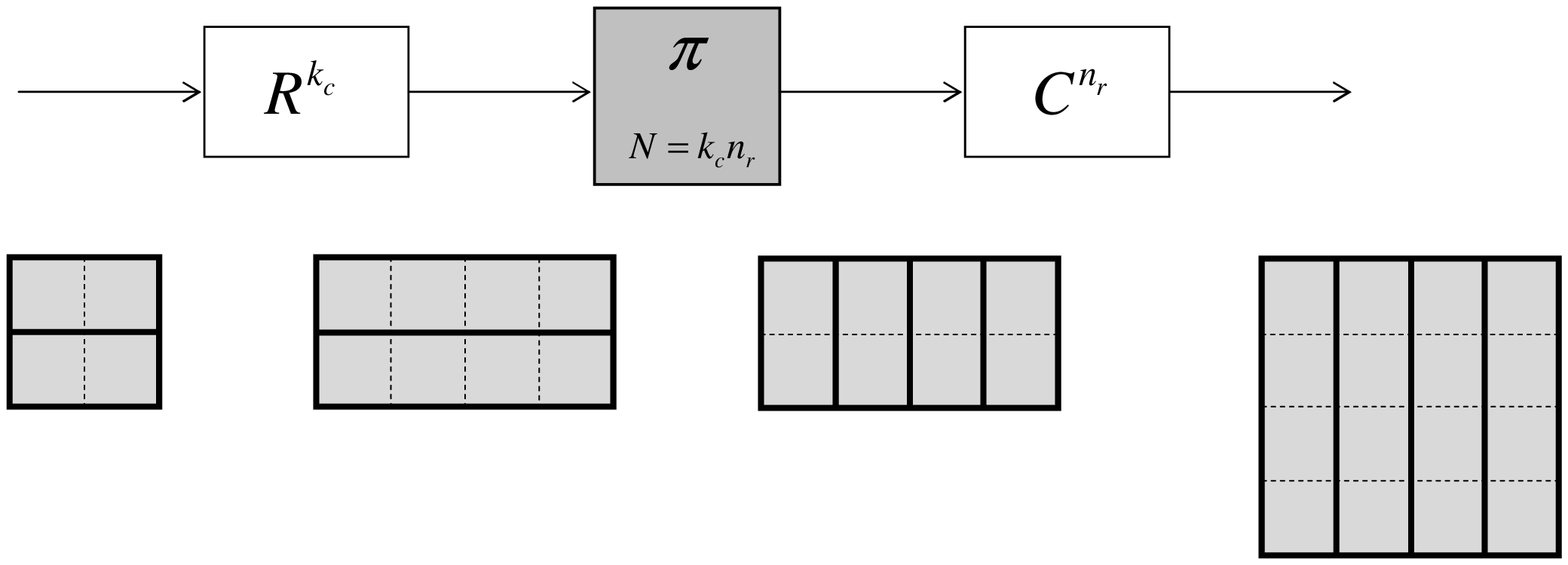}
\vspace*{2cm}
 \caption{Construction 1: serial concatenation.}
\label{fig:cos1}
\end{figure}

\clearpage

\begin{figure}
\centering
\includegraphics[height=3in, width=4in]{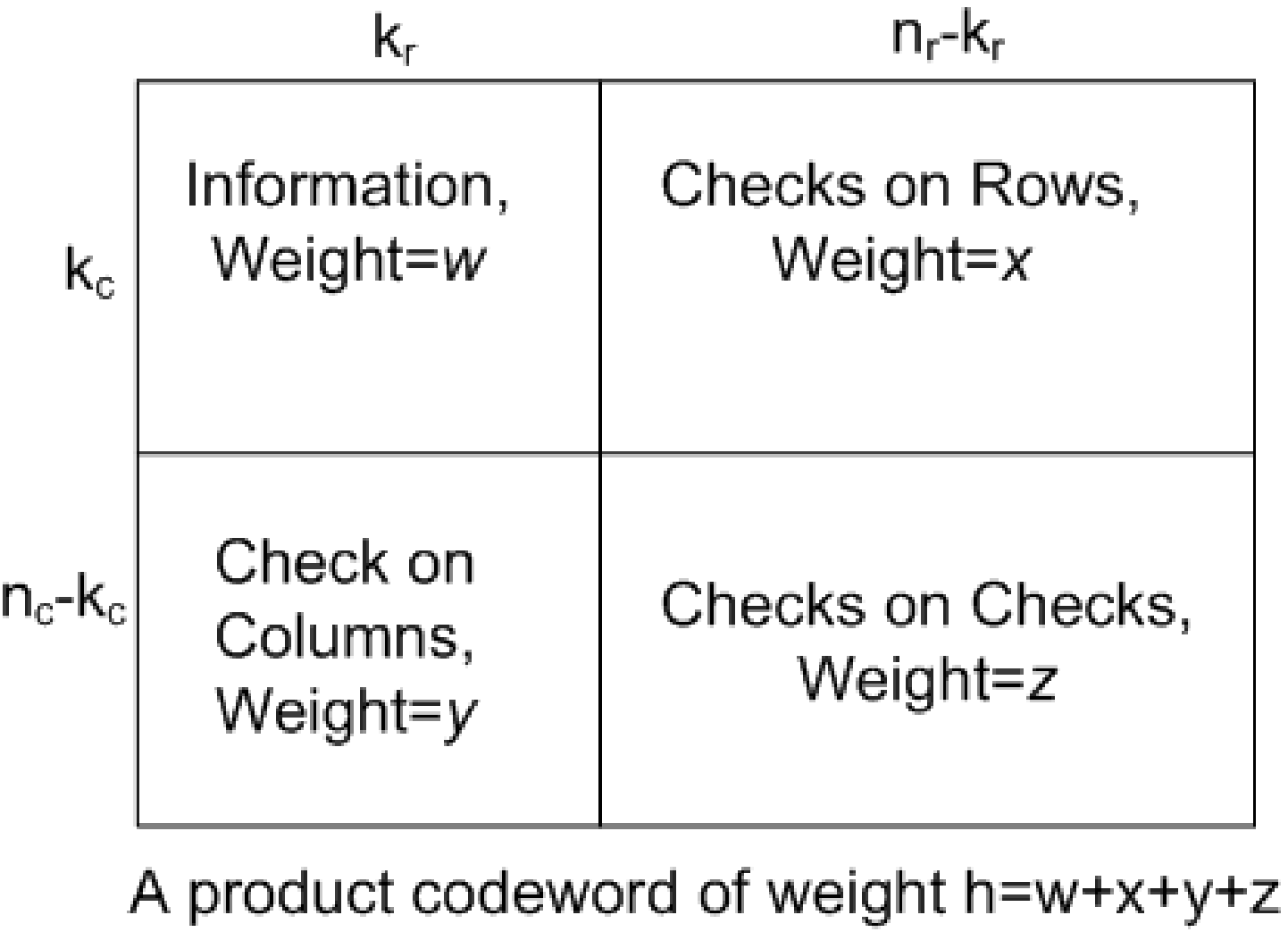}
\vspace*{2cm} \caption{\label{PC} The four set  partition of the
coordinates of a product codeword used in Construction 2.}
\end{figure}

\clearpage

\begin{figure}
\centering
\includegraphics[clip,scale=0.7,angle=270]{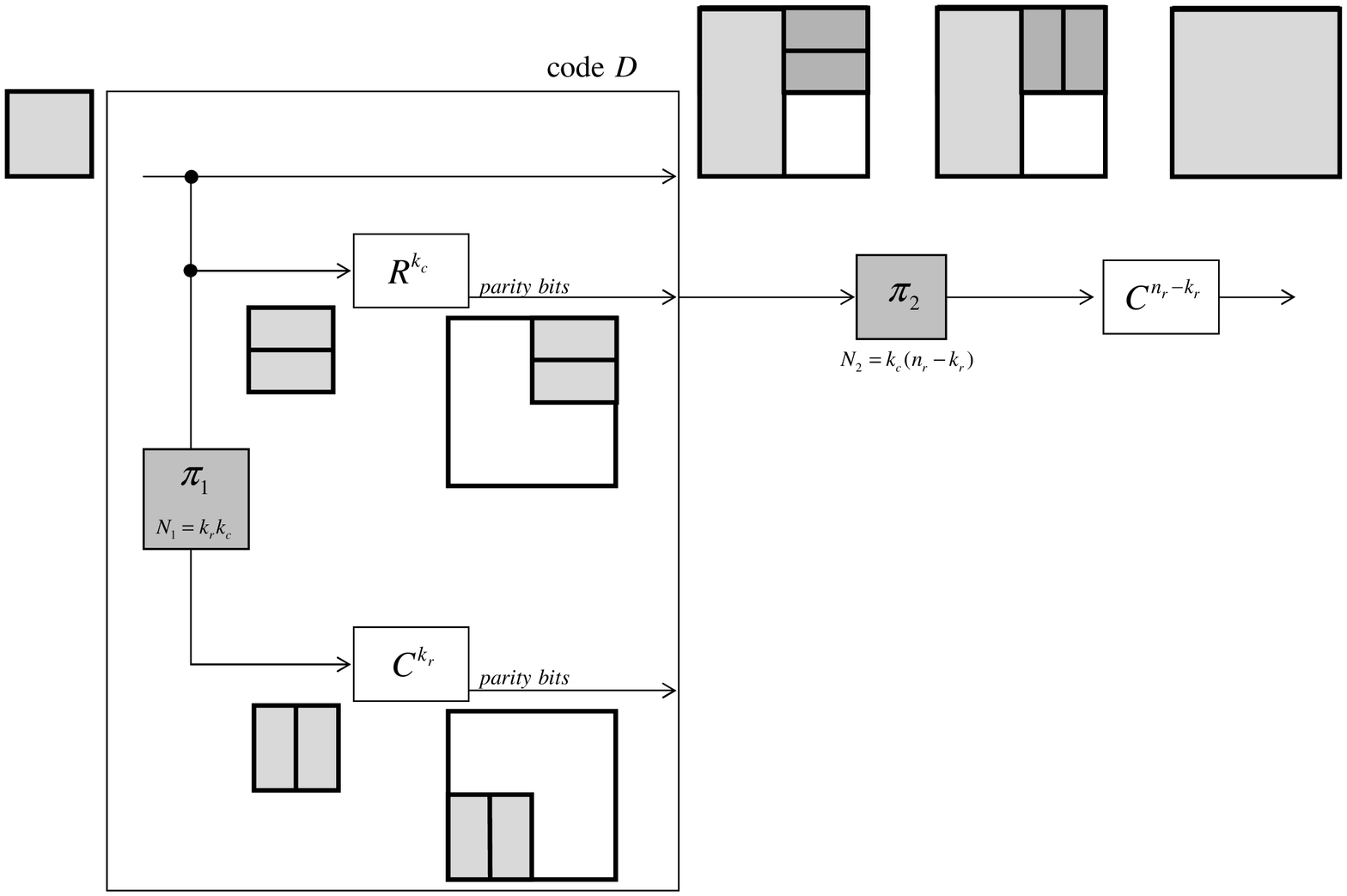}
\vspace*{2cm} \caption{Construction 2: parallel concatenation.}
\label{fig:cos2}
\end{figure}

\clearpage

\begin{figure}
\centering
\includegraphics[height=4in, width=5in]{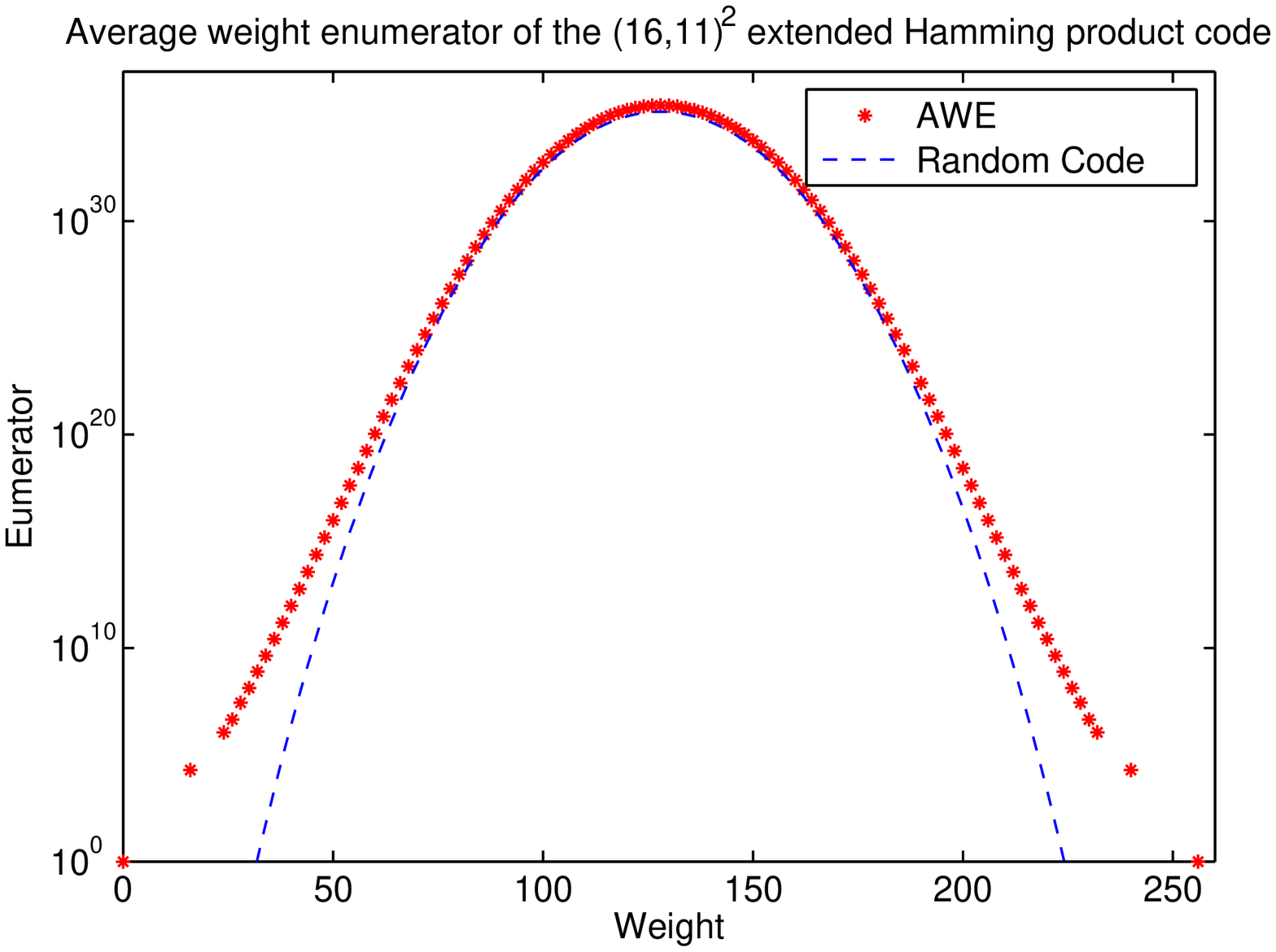}
\caption{\label{AWE1611EH} The combined weight enumerator of the
$(16,11)^2$ extended Hamming product code is compared with that of
a random binary code of the same dimension.}
\end{figure}

\clearpage

\begin{figure}
\centering
\includegraphics[height=4in, width=5in]{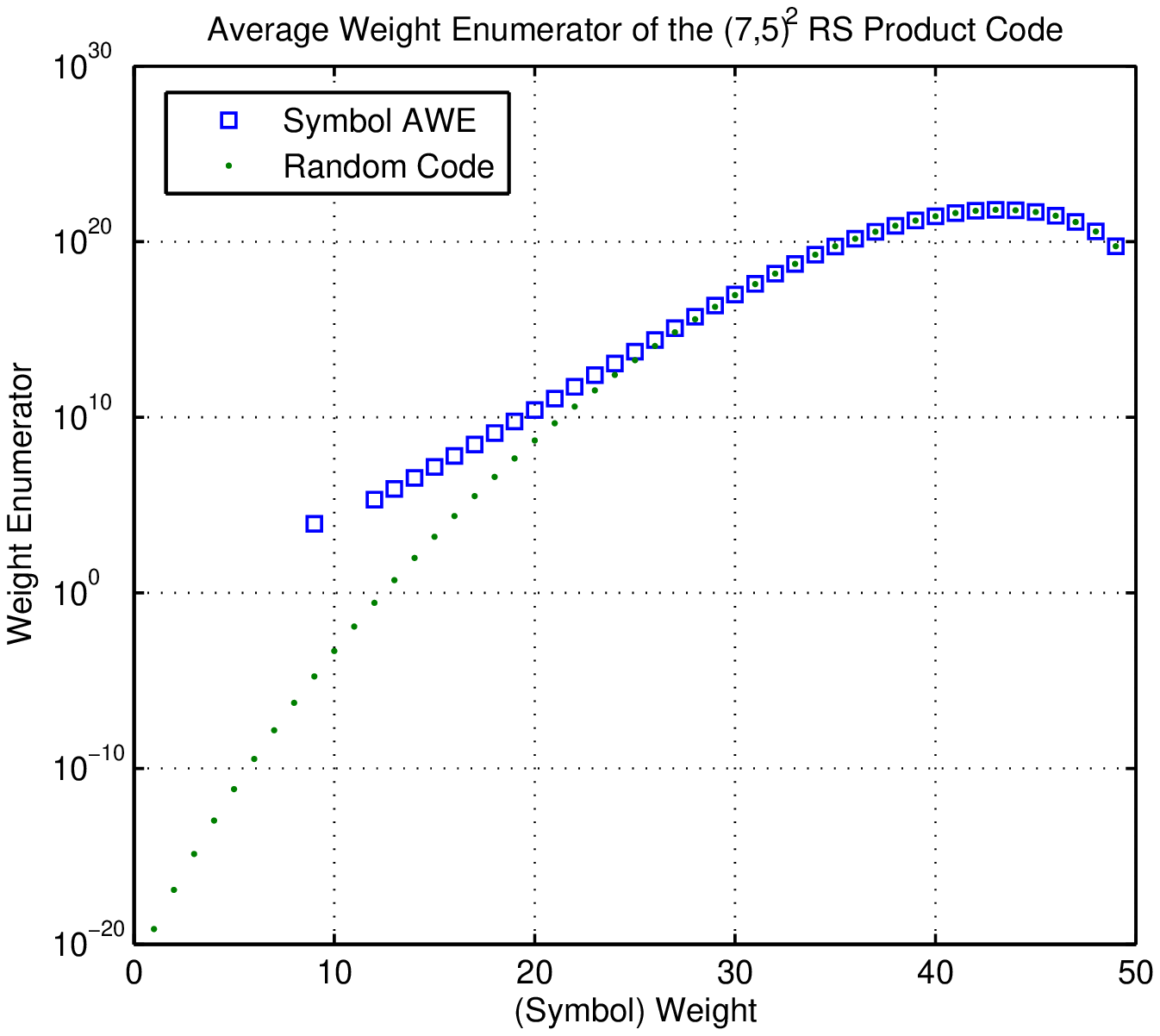}
\caption{\label{fRS75} The combined symbol weight enumerator of
the $(7,5)^2$ Reed Solomon product code is compared  with that of
a random code over $F_8$ with the same dimension. }
\end{figure}

\begin{figure}
\centering
\includegraphics[height=4in, width=5in]{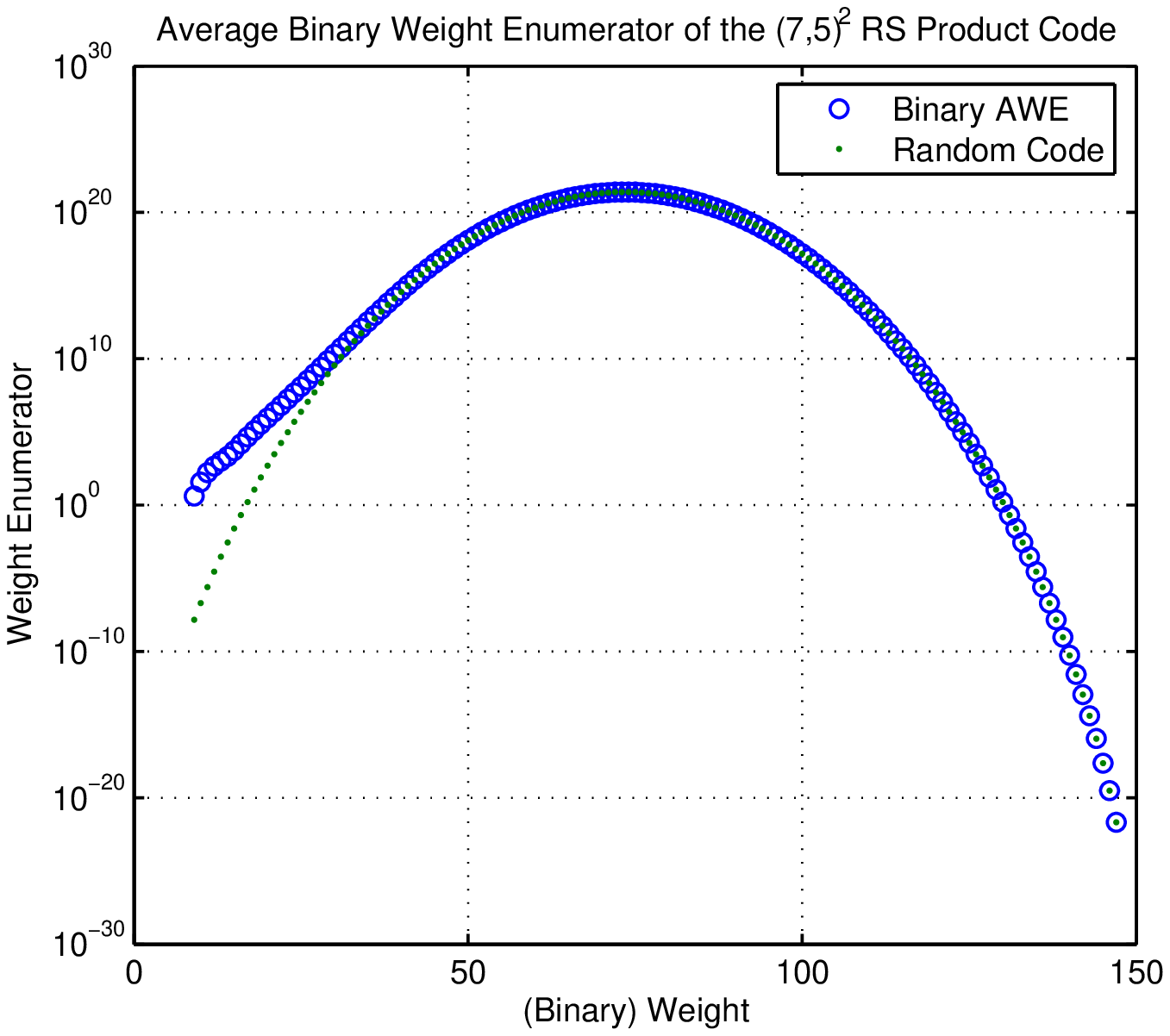}
\caption{\label{fRS752} The combined binary weight enumerator of
the binary image of the $(7,5)^2$ Reed Solomon product codes is
compared with that of a random binary code with the same
dimension.}
\end{figure}

\newpage

\begin{figure}
\centering
\includegraphics[height=6in,width=7in]{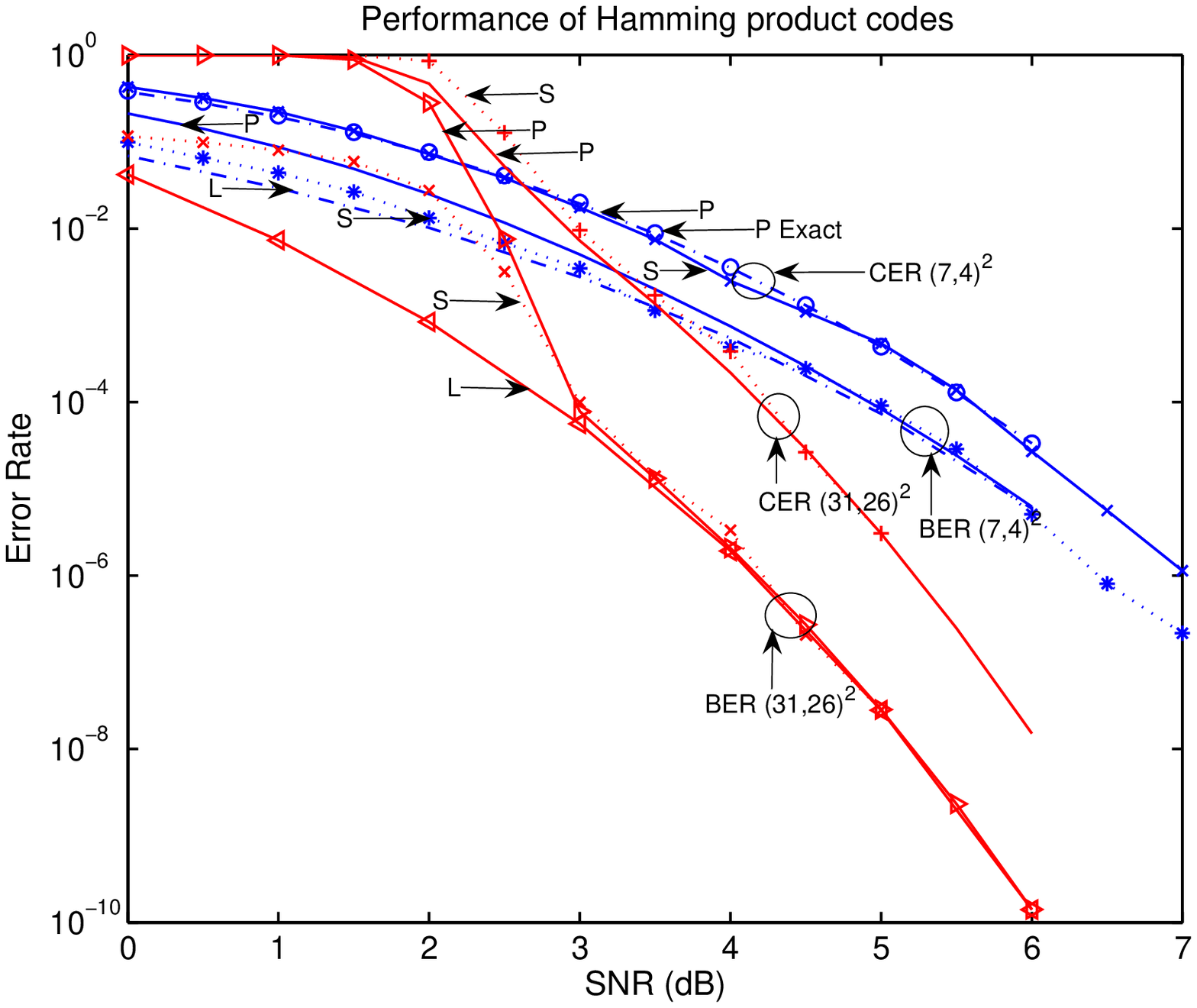}
\caption{\label{hamperf} CER and BER performance of some Hamming
product codes for soft decoding over AWGN channel.
 The Poltyrev bound P,
and the truncated union bound approximation L, are compared to
simulated performance of iterative decoding S. For the $(7,4)^2$
code, the Poltyrev bound computed with the exact weight enumerator
is also reported.
}
\end{figure}


\newpage

\begin{figure}
\centering
\includegraphics[height=4in,width=5in]{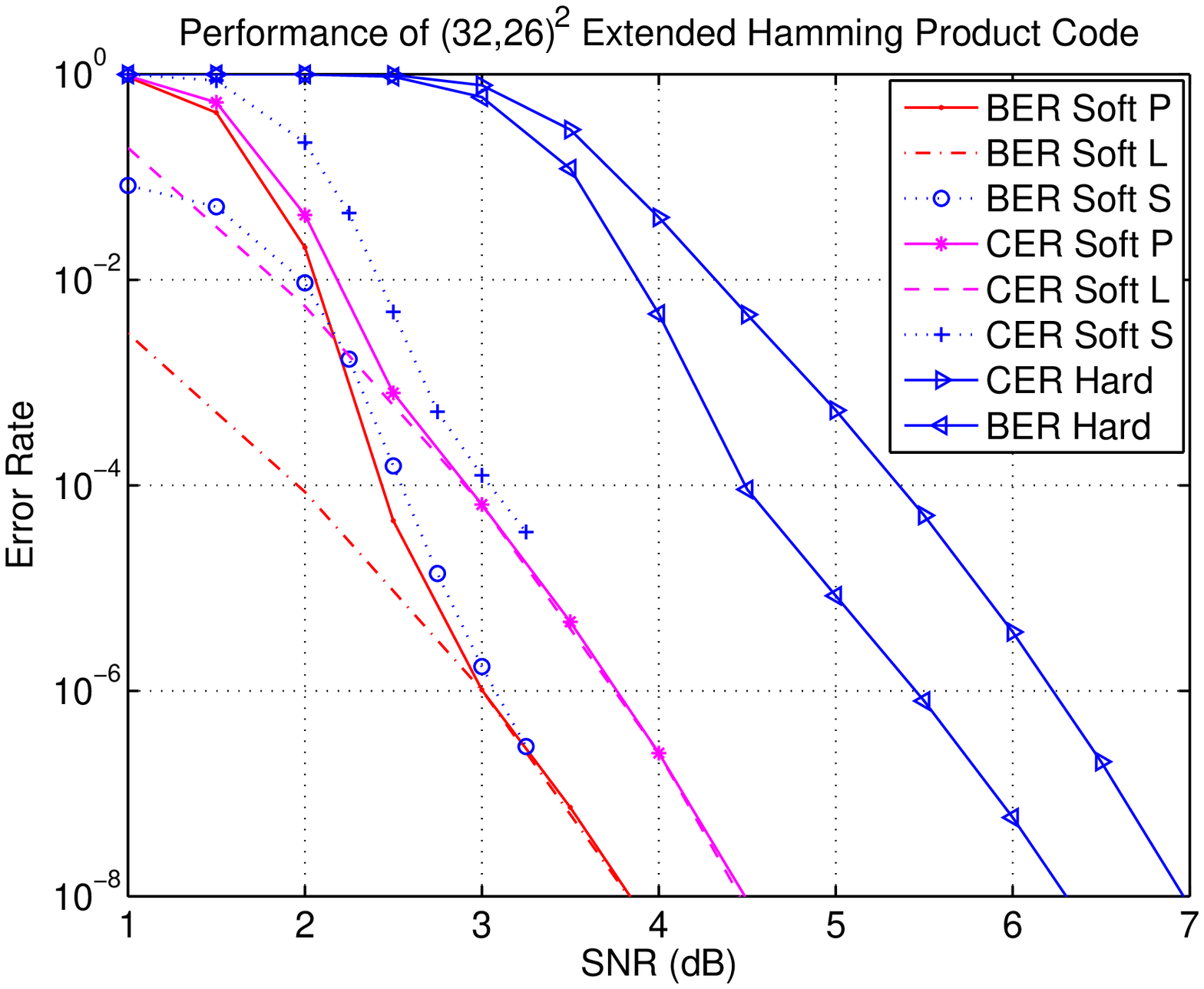}
\caption{\label{fEH32} CER and BER performance of the $(32,26)^2$
extended Hamming product code
for Soft and Hard decoding over AWGN channel.
%
The Poltyrev bound P and the truncated union bound approximation L
are compared to simulated performance of iterative  decoding S.}
\end{figure}
\newpage
 \begin{figure}
\centering

\includegraphics[height=3.5in,width=4.5in]{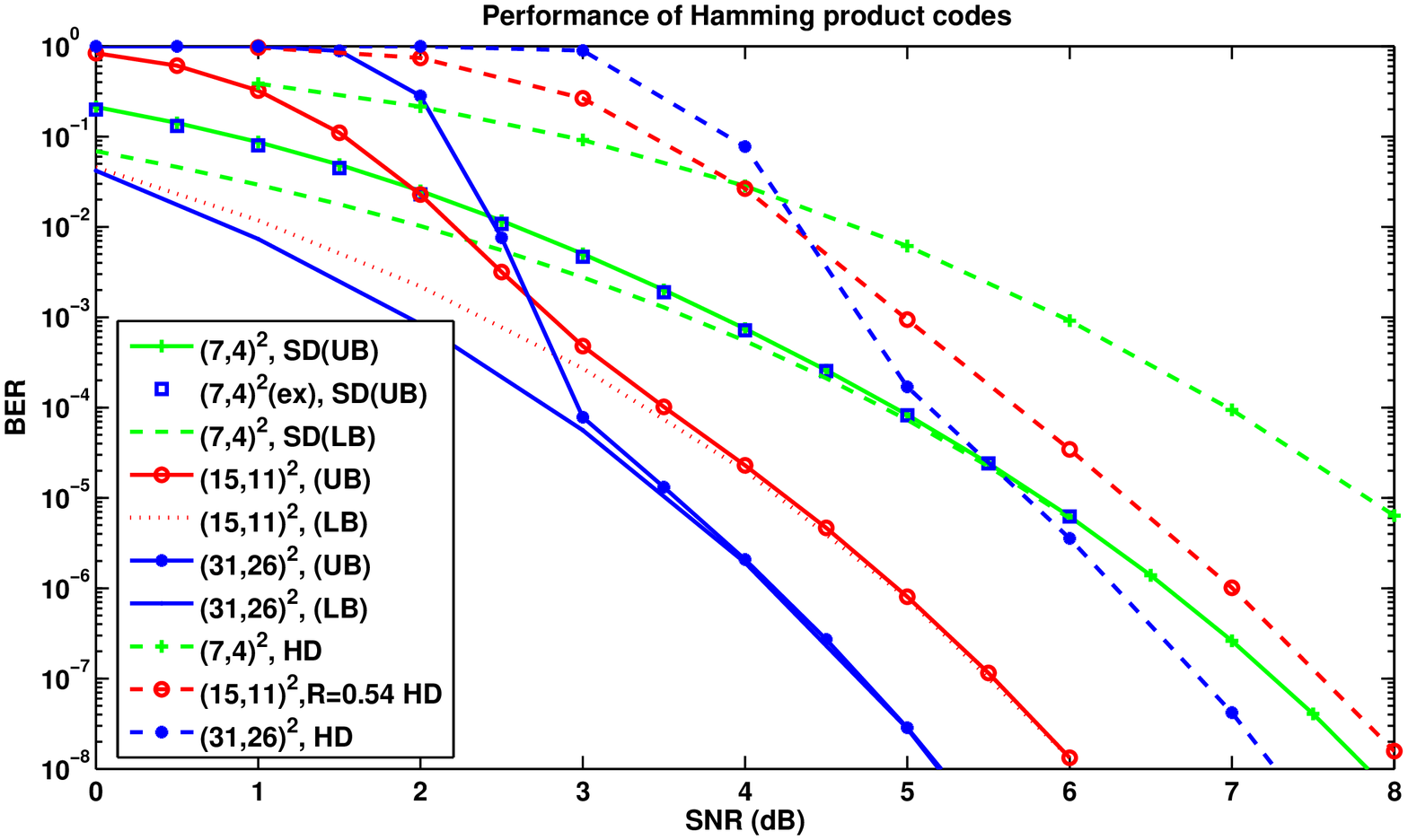}

\includegraphics[height=3.5in,width=4.5in]{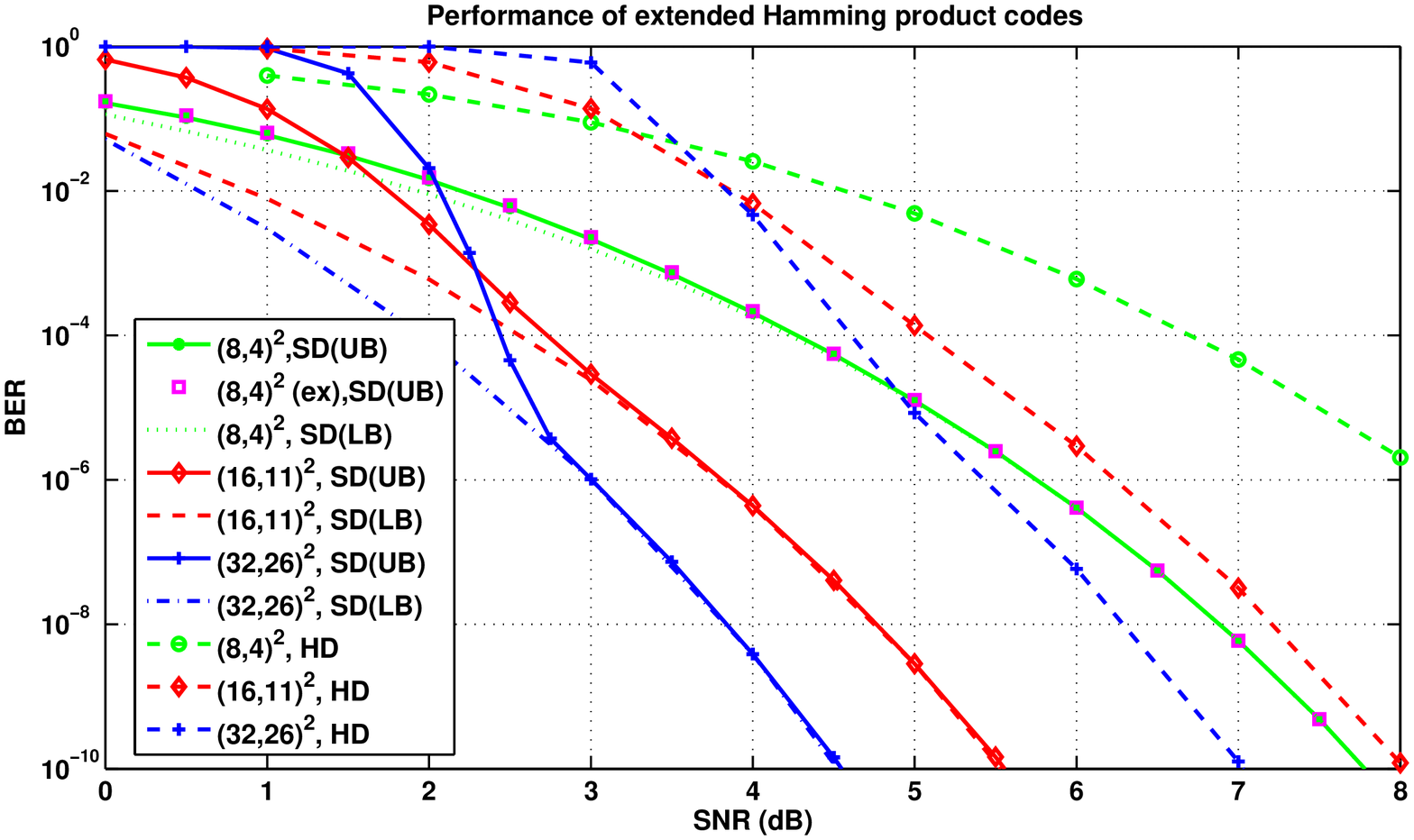}
\caption{\label{fEHfH} BER performance of  Hamming product codes
(upper) and extended Hamming product codes (lower) over AWGN
channel for both soft decision (SD) and (HD) hard decision. The
Poltyrev upper bound (UB) and the truncated union bound
approximation (LB) are used for SD, while the Poltyrev bound for
the BSC is used for HD.}
\end{figure}

\begin{figure}
\centering
\includegraphics[height=4in,width=5in]{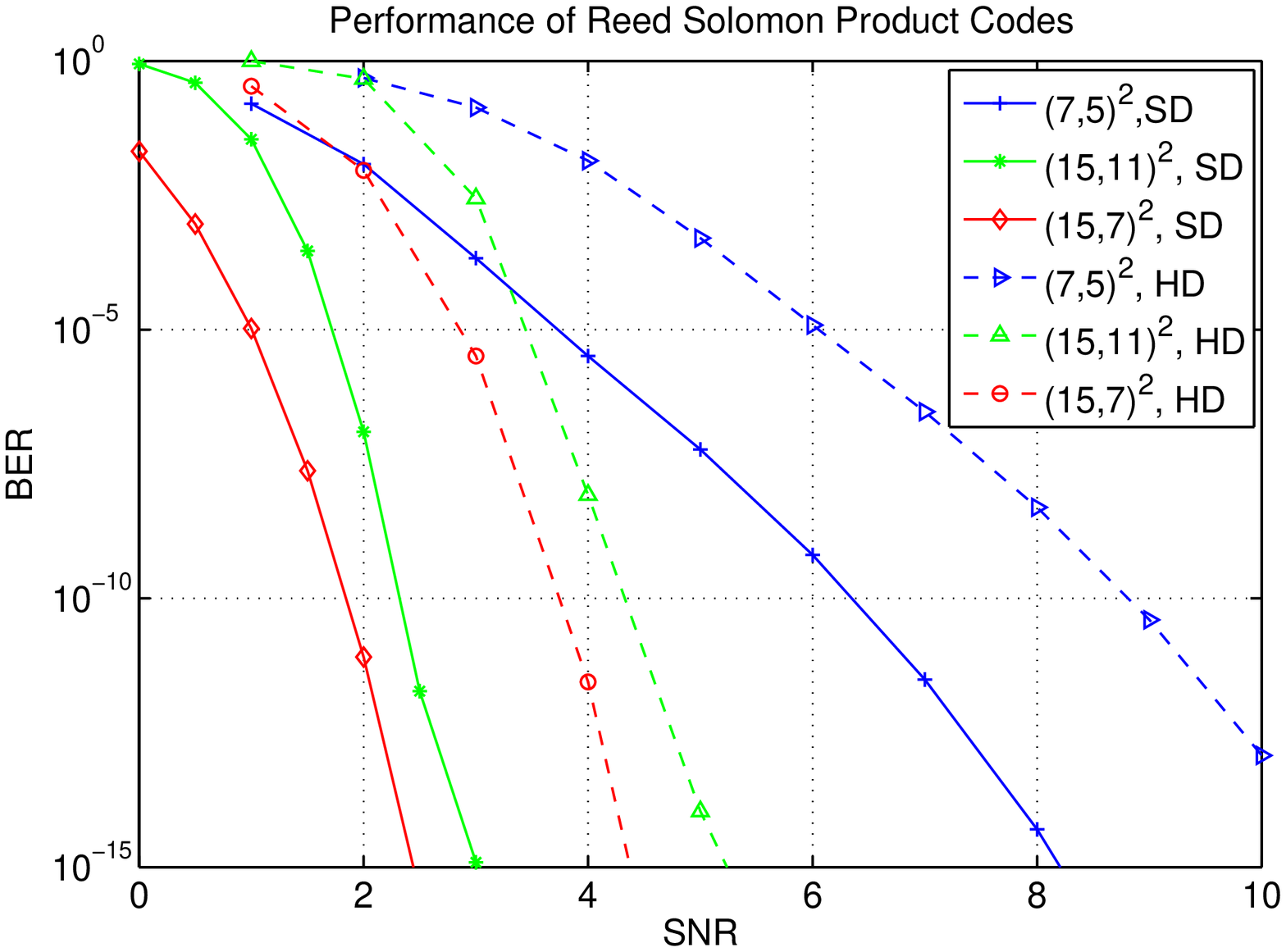}
\caption{\label{fRS} BER  performance of some Reed Solomon product
codes over the  AWGN  channel  for both soft decision (SD) and
hard decision (HD) decoding, obtained by plotting the Poltyrev
bound computed via the combined weight enumerators.
}
\end{figure}

\end{document}